\documentclass[aps,twocolumn,gbroupedaddress,reprint,amsmath,amssymb]{revtex4-2}
\usepackage[utf8]{inputenc}
\usepackage[T1]{fontenc}
\usepackage{amsmath}
\usepackage{amsfonts}
\usepackage{amssymb}
\usepackage{graphics,graphicx}
\usepackage{graphicx}
\usepackage{subfigure} %colocar figura lado a lado, figura a) e b).
\usepackage{wrapfig} % pacote reponsavel para colocar figura ao lado do texto
\usepackage{epstopdf}
\usepackage{xcolor}
\graphicspath{{figuras/}}

\usepackage{ulem}
\usepackage[breaklinks=true]{hyperref}
\usepackage{setspace}
\usepackage{graphicx}
\usepackage{color}

\begin{document}
      \title[]{Quasinormal modes, greybody factors and thermodynamics of four dimensional AdS black holes in Critical Gravity}

     \author{Jianhui Lin\textsuperscript{1}} 
     \author{Mois\'es Bravo-Gaete\textsuperscript{2}}\email{mbravo@ucm.cl}
     \author{Xiangdong Zhang\textsuperscript{1}}\email{Corresponding author. scxdzhang@scut.edu.cn}

     \address{\textsuperscript{1}Department of Physics, South China University of Technology, Guangzhou 510641, China, 
   	\\\textsuperscript{2}Departamento de Matem\'atica, F\'isica y Estad\'istica, Facultad de Ciencias B\'asicas, Universidad Cat\'olica del Maule, Casilla 617, Talca, Chile.}

\begin{abstract}
In the present work, considering critical gravity as a gravity model, an electrically charged topological Anti-de Sitter black hole with a matter source characterized by a nonlinear electrodynamics framework is obtained. This configuration is defined by an integration constant, three key structural constants, and a constant that represents the topology of the event horizon. Additionally, based on the Wald formalism, we probe that this configuration enjoys non-trivial thermodynamic quantities, establishing the corresponding first law of black hole thermodynamics, as well as local stability under thermal and electrical fluctuations. Additionally, via the Gibbs free energy we note that the topology of the base manifold allows us to compare this charged configuration with respect to the thermal AdS space-time, allowing us to obtain a first-order phase transition. The quasinormal modes and the greybody factor are also calculated by considering the spherical situation. We found that the quasinormal modes exhibit a straightforward change for variations of one of the structural constants.
\end{abstract}
\maketitle

%%%%%%%%%%%%%%%%%%%%%%
\section{Introduction}
%%%%%%%%%%%%%%%%%%%%%%

Analog to the ringing of a bell, when a black hole (BH) is perturbed, due for example to an infalling object or some disturbance in its surroundings, it responds via certain characteristic modes, denoted as quasinormal modes (QNMs) and the associated quasinormal frequencies (QNFs), that govern the time evolution of the initial perturbation \cite{Cardoso:2003pj}. These QNMs have attracted a high interest due to the detection of gravitational waves \cite{TheLIGOScientific:2017qsa,Monitor:2017mdv} (for reviews, see Refs. \cite{Kokkotas:1999bd,Berti:2009kk,Konoplya:2011qq}).    As was shown firstly by Regge and Wheeler \cite{Regge:1957td}, as well as by Vishveshwara \cite{Vishveshwara:1970zz}, at the moment of exploring Schwarzschild BHs, these modes depend only on parameters of these configurations, remaining independent of the nature of the perturbation and being a distinctive 'fingerprint' of the BH. The above has allowed the exploration of QNMs considering massless topological BHs \cite{Aros:2002te}, topological AdS BHs \cite{Koutsoumbas:2006xj,Gomez-Navarro:2017fyx}, regular rotating BHs \cite{Jusufi:2020odz} and with non-standard asymptotic behavior \cite{Becar:2015kpa}.

The consistent emergence and prominence of these distinctive perturbations have undergone rigorous testing, allowing us to study QNMs and QNFs deeply. For example, at linearized analyses \cite{Vishveshwara:1970zz,Davis:1971gg,Davis:1972ud}, where the fields are approached as perturbations in the singular BH spacetime,  numerical computations encompassing scenarios such as BH collisions \cite{Anninos:1993zj,Gleiser:1996yc} or stellar collapse \cite{Seidel,Smarr}, as well as the estimation of BH parameters, such as the mass, angular momentum, and charge, through gravitational waves \cite{Cardoso:2003pj,TheLIGOScientific:2017qsa,Monitor:2017mdv}. 

The evolution of small perturbations within BHs, generally unfolds in three key stages: the initial outburst, the ringing of QNMs, and the eventual power law tail. During the QNM ringing stage, the QNM from the BH shows damped oscillations, characterized by a discrete set of complex frequencies. Here, the real part represents the frequency of the oscillation, while the imaginary part indicates the rate at which the oscillation fades.

Concerning the calculation methodologies for QNMs, we will consider the finite element method as well as the pseudospectral methods \cite{Jansen:2017oag}. Together with the above, at the moment to consider quantum effect within a BH, the emission of Hawking radiation occurs \cite{Giddings:2015uzr}. Nevertheless, the presence of strong gravitational effects in its proximity acts as a barrier, implying that we observe is not a radiation spectrum, but rather a modified form known as a greybody spectrum. This spectrum is distinctly influenced by the curvature of spacetime, encapsulated by the greybody factor \cite{Boonserm:2023oyt,Visser:1998ke}.

Focused on the Anti-de Sitter/Conformal Field Theory (AdS/CFT) correspondence \cite{Maldacena:1997re,Gubser:1998bc,Witten:1998qj,Aharony:1999ti},  a large static BH asymptotically AdS (and its perturbation) corresponds to a thermal state in the CFT (perturbing this thermal state). Here, the QNFs in AdS BHs allow us to obtain a prediction for the thermalization timescale for a strongly coupled CFT \cite{Cardoso:2003pj}. In addition, from a thermodynamic point of view, the study of phase transitions (PTs) in strongly coupled field theories is one of the most relevant aspects of the gauge/gravity correspondence, where Hawking and Page, in their pioneering work \cite{Hawking-Page}, showed a PT between spherically symmetric AdS BHs and thermal AdS space-time, this represents confining and deconfining PTs in the dual quark-gluon plasma \cite{Witten:1998qj,Witten:1998zw}. Along with the above, PTs have been an object of high study when considering AdS-charged BHs, which show a resemblance to the Van der Waals fluid \cite{Chamblin:1999tk,Chamblin:1999hg}.

On the other hand, by exploring gravity theories beyond General Relativity, L\"{u} and Pope in \cite{Lu:2011zk} showed that via the action
\begin{eqnarray}\label{eq:CG}
		S_{\tiny{\mbox{CG}}}[g_{\mu \nu}, R_{\mu \nu \sigma \rho}]&=& \int{d}^4x\sqrt{-g}\mathcal{L}_{\tiny{\mbox{CG}}} \nonumber\\
                          &=&\int{d}^4x\sqrt{-g} \left[\frac{R-2\Lambda}{2 \kappa} \right. \\
 &+&\left.\frac{1}{2 \kappa}\left(-\frac{1}{2 \Lambda} {R}^2+\frac{3}{2 \Lambda} {R}_{\alpha\beta}{R}^{\alpha\beta} \right)\right], \nonumber
\end{eqnarray}
where $\kappa$ is a coupling constant and $\Lambda$ the cosmological constant, a four-dimensional renormalizable theory devoid of ghosts appears, being known as Critical Gravity (CG). For the theory (\ref{eq:CG}),  the massive scalar mode is eliminated, while the massive spin-2 particle becomes massless,  and the on-shell energy of the remaining massless gravitons becomes zero. Looking at it from a thermodynamic perspective, and according to the authors, although the theory (\ref{eq:CG}) supports the AdS metric as well as the well-known Schwarzschild AdS (Sch-AdS) BH,  the adjustments integrated into the parameters of the action (\ref{eq:CG}) yield null thermodynamic extensive quantities, being the price to pay to obtain a well-behaved theory \cite{Lu:2011zk}. The above result has generated the exploration of BHs, supported by CG (\ref{eq:CG}), with non-vanishing thermodynamic properties. One of them is via with non-standard asymptotic behavior \cite{Bravo-Gaete:2021kgt}, through the inclusion of dilatonic fields and a non-minimally coupled scalar field, or asymptotically AdS configurations with planar base manifold \cite{Alvarez:2022upr,Paul:2023vys} with nonlinear electrodynamics (NLE). For instance, NLE can help circumvent the singularity of the field of a point particle in standard electrodynamics, attempt to describe quantum electrodynamics \cite{Cembranos:2014hwa}, and obtain stationary solutions \cite{DiazGarcia:2022jpc,Garcia-Diaz:2021bao,Ayon-Beato:2022dwg}. This has given rise to well-known theories such as the Born and Infeld theory \cite{Born:1934,Born.Infeld:1934,Born.Infeld:1935,Weiss:1937jtb,Infeld:1936,Infeld:1937}, the Euler-Heisenberg models \cite{Heisenberg:1936nmg},  ModMax electrodynamics
\cite{Bandos:2020jsw,Kosyakov:2020wxv} and the construction of a NLE through a sum of infinite series of Maxwell invariant \cite{Gao:2021kvr}.

The search for new AdS configurations supported by CG with non-vanishing thermodynamic properties and motivated by gauge/gravity duality, has led to the consideration of NLE as a matter source. In the present work, we consider a nonlinear behavior of the electromagnetic field $A_\mu$ with field strength $F_{\mu \nu}=\partial_{\mu} A_{\nu}-\partial_{\nu} A_{\mu}$, and the antisymmetric tensor $\mathcal{P}_{\mu\nu}$ (known as Pleb\'anski tensor), which reads \cite{Plebanski:1968,Salazar:1987}:
\begin{eqnarray}\label{eq:NLE}
		S_{\tiny{\mbox{NLE}}}[A_{\mu}, P_{\mu \nu}]&=& \int{d}^4x\sqrt{-g}\mathcal{L}_{\tiny{\mbox{NLE}}}\\
		 &=&
		\int{d}^4x\sqrt{-g} \left(-\frac{1}{2}\mathcal{P}^{\mu\nu}F_{\mu\nu}+\mathcal{H}(\mathcal{P})\right). \nonumber
\end{eqnarray}
The introduction of  $\mathcal{P}_{\mu\nu}$ arises from the need to establish a relationship between a standard electromagnetic theory with Maxwell's theory of continuous media, together with a structure-function $\mathcal{H}(\mathcal{P})$ depending on the invariant formed with  $\mathcal{P}_{\mu\nu}$ ($\mathcal{P}: = \frac{1}{4}\mathcal{P}_{\mu\nu}\mathcal{P}^{\mu\nu}$) where the linear Maxwell scenario is naturally recovered when $\mathcal{H}(\mathcal{P}) \propto \mathcal{P}$. The action (\ref{eq:NLE}) has been studied
extensively in the context of gravitational theories, regular non-rotating BHs, and charged rotating BHs configurations (see Refs. \cite{Ayon-Beato:1998hmi,Ayon-Beato:1999qin,Ayon-Beato:1999kuh,Ayon-Beato:2000mjt,Ayon-Beato:2004ywd,Garcia-Diaz:2021bao,DiazGarcia:2022jpc}). 

Given the aforementioned details, in the present paper, our first objective is (i) to show that the planar extension of the work present in \cite{Alvarez:2022upr} can be extended to an arbitrary topology base manifold, allowing us to explore asymptotically AdS BHs, with one or even more locations of the event horizon. Additionally, via the Wald formalism as procedure \cite{Wald:1993nt, Iyer:1994ys}, we present that these charged BHs configurations enjoy non-zero thermodynamic properties, wherein the topology plays a crucial role. In fact, via Gibbs free energy and comparing with respect to the thermal AdS space-time, we will have first-order PTs as well as situations where these BHs are the preferred configuration. With this information, the second objective is (ii) to delve into the study of QNMs as well as the calculation of greybody factors,  considering the spherical situation.

The novelty of the present work is based on two key dimensions. Firstly, by combining the gravity theory (\ref{eq:CG}) with the matter source (\ref{eq:NLE}), it opens avenues to explore charged BHs configurations beyond the planar scenario, as well as their thermodynamic properties. Secondly, it introduces an opportunity to explore QNMs for the spherical case, where the influence of quadratic corrections stemming from CG (\ref{eq:CG}) and NLE as a matter source becomes a significant factor.

The structure of this paper unfolds in the following manner: In Section \ref{Sec-BH} we present the charged BH solution, exploring the existence of horizons independent of the topology of the event horizon, while in Section \ref{Sec-Termo} the thermodynamic properties are explored. In Section \ref{Sec-QNM} the QNMs and the greybody factor are computed and analyzed. Finally, Section \ref{Sec-conclusions} is devoted to our conclusions and discussions.

%%%%%%%%%%%%%%%%%%%%%%%%%%%%%%%%%%%%%%%%%%%%%%%%%%%%%%%
\section{The analysis of the solution}\label{Sec-BH}
%%%%%%%%%%%%%%%%%%%%%%%%%%%%%%%%%%%%%%%%%%%%%%%%%%%%%%%

To perform a complete exploration and analysis for this charged BH configurations, the equation of motions that result from the variation of the action 
\begin{eqnarray}\label{eq:totaction}
S[g_{\mu \nu}, R_{\mu \nu \sigma \rho},A_{\mu}, P_{\mu \nu}]&=&S_{\tiny{\mbox{CG}}}[g_{\mu \nu}, R_{\mu \nu \sigma \rho}]\\
&+&S_{\tiny{\mbox{NLE}}}[A_{\mu}, \mathcal{P}_{\mu \nu}],\nonumber
\end{eqnarray}
are required, which are given by
\begin{subequations}\label{eq:EOM}
\begin{eqnarray}
\mathcal{E}^{\nu}_{F}&=&\nabla_{\mu} P^{\mu\nu}=0, \label{eq:Maxwell}\\
\mathcal{E}^{\mu \nu}_{P} &=&-F^{\mu\nu}+\left(\frac{\partial \mathcal{H}}{\partial \mathcal{P}}\right) \mathcal{P}^{\mu\nu}=0,
%\textrm{where} \,\,
% \mathcal{H}_P\equiv \partial \mathcal{H}/\partial P.
\label{eq:constitutive} \\
\mathcal{E}_{\mu \nu}&=&{\mathcal{G}_{\mu\nu}^{CG}}-\kappa T_{\mu\nu}^{NLE}=0,
\label{eq:Einstein}
\end{eqnarray}
 and the tensors ${\mathcal{G}_{\mu\nu}^{CG}}$ and $T^{NLE}_{\mu\nu}$ are defined as follows:
\begin{eqnarray}
{\mathcal{G}_{\mu\nu}^{CG}}&=&G_{\mu \nu}+\Lambda g_{\mu \nu}+
\frac{3}{\Lambda} \Bigl(R_{\mu\rho}R_{\nu}^{\, \rho}
 -\frac{1}{4}R^{\rho\sigma}R_{\rho\sigma}g_{\mu\nu}\Bigr)\nonumber \\
&-& \frac{1}{\Lambda}  \Bigl( R R_{\mu\nu}-\frac{1}{4} R^2 g_{\mu\nu}\Bigr)
+\frac{3}{2 \Lambda}  \Bigl(\Box R_{\mu\nu} + \frac{1}{2} \Box R g_{\mu\nu}\nonumber\\
&-& 2 \nabla_{\rho}\nabla_{\left(\mu\right.}R_{\left.\nu\right)}^{\,  \rho} \Bigr)
-\frac{1}{\Lambda} (g_{\mu\nu}\Box R-\nabla_{\mu}\nabla_{\nu} R), \nonumber\\
T_{\mu\nu}^{NLE}&=&  \left(\frac{\partial \mathcal{H}}{\partial \mathcal{P}}\right) \mathcal{P}_{\mu\alpha}\mathcal{P}_{\nu}^{\, \alpha}
-g_{\mu\nu}\left(2\mathcal{P} \left(\frac{\partial \mathcal{H}}{\partial \mathcal{P}}\right)-\mathcal{H}\right).\nonumber
\end{eqnarray}
\end{subequations}
Here,  eq. (\ref{eq:Maxwell}) stands as the nonlinear rendition of Maxwell's equations, encapsulating the constitutive relations within (\ref{eq:constitutive}). Meanwhile,  (\ref{eq:Einstein}) represents the Einstein equations. For the  model  (\ref{eq:CG})-(\ref{eq:totaction}), we consider the following four-dimensional metric Ansatz:
\begin{eqnarray}
ds^{2}&=&-\left(\epsilon+\frac{r^2}{l^2} f(r)\right)dt^{2}+\left(\epsilon+\frac{r^2}{l^2} f(r)\right)^{-1}{dr^{2}}
\nonumber\\
&+&r^{2}d\Omega^{2}_{2,\epsilon},\label{eq:metric4d}
\end{eqnarray}
where $t \in \,(-\infty,+\infty), r > 0$, $l$ is the AdS radius, and the metric function must satisfy the asymptotic condition $${\lim_{r\rightarrow +\infty}f(r) = 1,}$$ 
while that $d\Omega^{2}_{2,\epsilon}$ represents the line element given by
\begin{eqnarray}\label{eq:dOmega2}
d\Omega^{2}_{2,\epsilon} = \left\{ \begin{array}{ll}
d\theta^{2}+\sin^{2}(\theta) d\rho^{2}, & \mbox{for $ \epsilon=1,$}\\
d\theta^{2}+\sinh^{2}(\theta) d\rho^{2}, & \mbox{for $ \epsilon=-1,$}\\
d\theta^{2}+d \rho^{2}, & \mbox{for $ \epsilon=0,$}
\end{array} \right. 
\end{eqnarray}
indexed through the constant $\epsilon$. For our analysis, we consider purely electrical configurations; this is
$\mathcal{P}_{\mu\nu}=2\delta^t_{[\mu}\delta_{\nu]}^r G(r)$ where,  if we replace it in the nonlinear Maxwell equation (\ref{eq:Maxwell}), we obtain
$$\mathcal{P}_{\mu\nu}=2\delta_{[\mu}^t\delta_{\nu]}^r\frac{M}{r^2},$$
where $M$ is an integration constant. Therefore, the electric invariant $\mathcal{P}$ is negative definite since we only consider purely electrical configurations, which reads
\begin{equation}\label{eq:P}
\mathcal{P}=\frac{1}{2} {\mathcal{P}_{rt}\mathcal{P}^{rt}}=-\frac{M^2}{2r^4}<0.
\end{equation}
With the above information, we note that the difference between the temporal and radial diagonal components of Einstein's equations (\ref{eq:Einstein}) (this is $ \mathcal{E}_{t}^{t}-\mathcal{E}_{r}^{r}=0$) is proportional to a fourth-order Cauchy-Euler ordinary differential equation, where the metric function $f$ enjoys the structure $f(r)=M_{0}+\sum_{i=1}^{3}M_{i}\left(\frac{l}{r}\right)^{i}$, with $M_i$'s integration constants, while that via the combination of $\mathcal{E}^{t}_{t}-\mathcal{E}_{\theta}^{\theta}=0$ (or $\mathcal{E}^{t}_{t}-\mathcal{E}_{\rho}^{\rho}=0$), and expressing the radial coordinate $r$ in terms of the 
electric invariant $\mathcal{P}$ from eq. (\ref{eq:P}), one can determine $\partial \mathcal{H}/ \partial \mathcal{P}$, which can be later integrated to obtain $\mathcal{H}(\mathcal{P})$, which reads \cite{Alvarez:2022upr}
\begin{eqnarray}\label{eq:H}
\mathcal{H}(\mathcal{P}) &=&  \frac{ (\alpha_2^2-3 \alpha_1 \alpha_3)l^2 \mathcal{P}}{3\kappa}
{-\frac{2\alpha_1 (-2 \mathcal{P})^{1/4}}{l\kappa}}\nonumber\\
&+&{\frac{\alpha_2\sqrt{-2 \mathcal{P}} }
{\kappa}},
\end{eqnarray}
where $\alpha_1$, $\alpha_2$ and $\alpha_3$ are coupling constants, together with a metric function
\begin{eqnarray}\label{eq:f}
f(r) &=&1-\alpha_1\sqrt{M} \left(\frac {l}{r}\right)+\alpha_2 M \left(\frac {l}{r}\right)^2\nonumber\\
&-&\sqrt{M} \left(\alpha_3 M+\frac{2 \epsilon \alpha_2}{3 \alpha_1}\right) \left(\frac {l}{r}\right)^3,
\end{eqnarray}
after a redefinition of the integration constants $M_i$'s. Finally, to satisfy  the equations of motion (\ref{eq:Maxwell})-(\ref{eq:Einstein}), the cosmological constant must be fixed as 
\begin{eqnarray}\label{eq:Lambda}
\Lambda=-\frac{3}{l^2},
\end{eqnarray}
 where we note that for the planar situation ($\epsilon=0$), the configuration obtained in \cite{Alvarez:2022upr} is naturally recovered. It is important to note that the uncharged Sch-AdS case can be obtained from the structure-function (\ref{eq:H}) as well as the metric function (\ref{eq:f}) by first fixing $\alpha_2=0$ and then $\alpha_1=0$. This implies that even the linear Maxwell case is not allowed, which emphasizes the importance of including this matter source.

\begin{figure}[h!]
  \centering
    \includegraphics[scale=0.22]{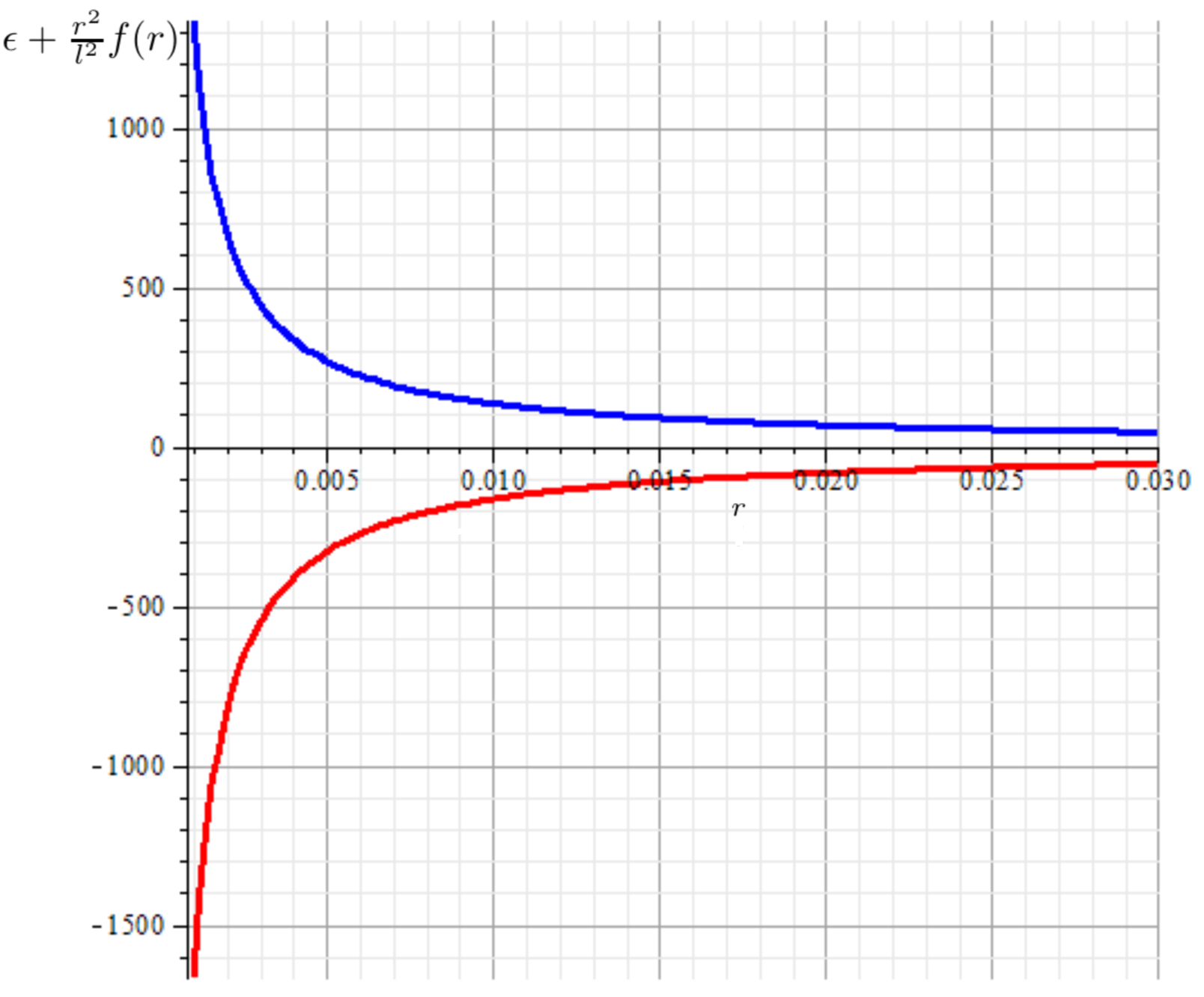}
      \caption{\label{fig1} Behavior of the metric function $\epsilon+\frac{r^2}{l^2} f(r)$ when $r \rightarrow 0^{+}$. In this context, the red curve represents Case 1, while the blue curve represents Case 2. For our calculations, we consider the values for the blue curve  $\alpha_1=\alpha_2=1,\alpha_3=-2,M=1,\epsilon=1, l=1$, while that for the red curve  $\alpha_1=\alpha_2=\alpha_3=1,M=1,\epsilon=1,l=1$.}
\end{figure}

On the other hand, regarding the line element (\ref{eq:metric4d}) and the metric function (\ref{eq:f}), in order to obtain a BH configuration, some conditions need to be established. First that all, the scalar curvature $R$ reads
\begin{eqnarray}\label{eq:R}
	R&=& -\frac{12f}{l^2}-\frac{8 rf'}{l^2}-\frac{r^2 f''}{l^2},\nonumber\\
	&=&-\frac{12}{l^2}+\frac{6 \alpha_1 \sqrt{M}}{rl}-\frac{2 \alpha_2M}{r^2},
\end{eqnarray}
where for our notations ($')$ denotes the derivative with respect to coordinate $r$, and showing us a curvature singularity located at $r_s=0$. With respect to the existence of event horizon, this is $r_h>0$ such that 
$$\Big(\epsilon+\frac{r^2}{l^2} f(r)\Big) \Big{|}_{r=r_h}=0, \mbox{ and } \Big(\epsilon+\frac{r^2}{l^2} f(r)\Big)' \Big{|}_{r=r_h}>0,$$ we note that in the limit, as $r$ approaches positive infinity, we have that $\epsilon+\frac{r^2}{l^2} f(r) \simeq \frac{r^2}{l^2}$.  While that when $r \rightarrow 0^{+}$,  $\epsilon+\frac{r^2}{l^2} f(r) \simeq -\sqrt{M} \left(\alpha_3 M+\frac{2 \epsilon \alpha_2}{3 \alpha_1}\right) \frac {l}{r}$, allowing us to split the analysis in two cases:
\begin{eqnarray}
\mbox{{\bf{Case 1:}}} {\mbox{ If }}  \alpha_3 M+\frac{2 \epsilon \alpha_2}{3 \alpha_1}>0  &:& \epsilon+\frac{r^2}{l^2} f(r)  \mbox{ starts} \nonumber\\
& & \mbox{increasing from } \nonumber\\
& &\mbox{$-\infty$ when } \nonumber\\
& &\mbox{$r$ increases.} 
\label{eq:Cases}
\end{eqnarray}
\begin{eqnarray}
\mbox{{\bf{Case 2:}}} {\mbox{ If }}  \alpha_3 M+\frac{2 \epsilon \alpha_2}{3 \alpha_1}<0  &:& \epsilon+\frac{r^2}{l^2} f(r)  \mbox{ starts}\nonumber\\
& & \mbox{decreasing from } \nonumber\\
& &\mbox{$+\infty$ when} \nonumber\\
& &\mbox{$r$ increases,} \nonumber
\end{eqnarray} 
as is shown in Figure \ref{fig1}.

\begin{figure}[h!]
  \centering
    \includegraphics[scale=0.23]{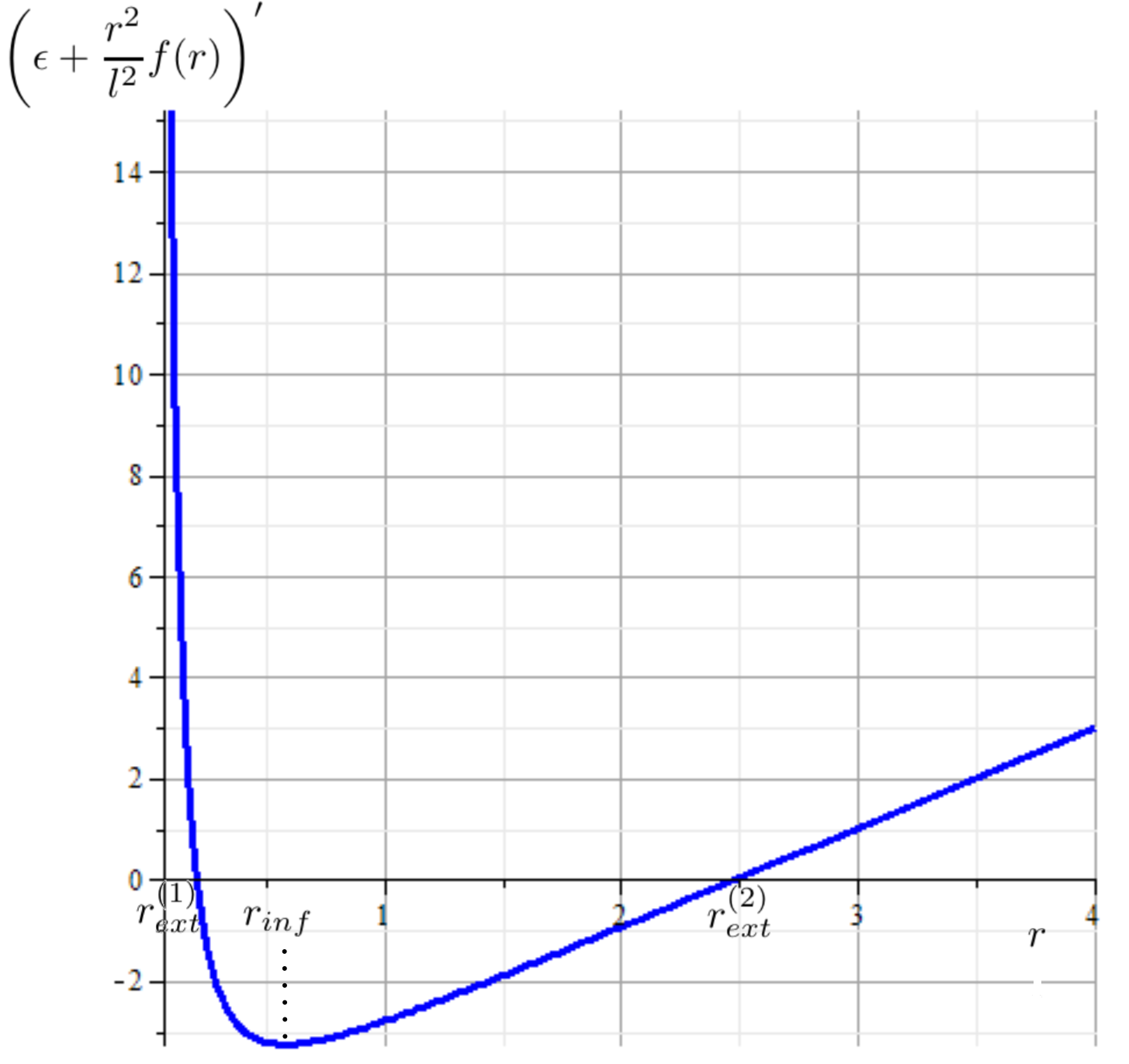}
      \caption{\label{fig2} Behavior of  $\left(\epsilon+\frac{r^2}{l^2} f(r)\right)'$ for Case 1 when the inequalities (\ref{eq:in1}) are satisfied. In this context, the existence of a maximum ($r_{\text{ext}^{(1)}}$) point, inflection point ($r_{\text{inf}}$) and minimum point ($r_{\text{ext}^{(2)}}$) is ensured, when  $r_{\text{ext}^{(1)}}<r_{\text{inf}}<r_{\text{ext}^{(2)}}$. For our computations, we consider  $M=1,\alpha_1=5,\alpha_2=1/4,\alpha_3=1/6,\epsilon=l=1$.}
\end{figure}

For these situations, we can conduct a study of the extreme values, denoted as $r_{\text{ext}}>0$, for  $\epsilon+\frac{r^2}{l^2} f(r)$, via the first derivative with respect to the radial coordinate $r$. For this analysis, we observe that these values satisfy
\begin{eqnarray}
&&\Big(\epsilon+\frac{r^2}{l^2} f(r)\Big)'{\Big{|}}_{r=r_{\text{ext}}} \label{eq:ext} \\
&&=\Big[ 2\Big(\frac{r}{l}\Big)-\alpha_1\sqrt{M} +\frac{l^2}{r^2} \sqrt{M} \Big(\alpha_3 M+\frac{2 \epsilon \alpha_2}{3 \alpha_1}\Big)\Big]{\Big{|}}_{r=r_{\text{ext}}}\nonumber\\
&&=0.\nonumber
\end{eqnarray}
Furthermore, to determine whether these extreme values correspond to a maximum or minimum, we examine the sign of the expression
\begin{eqnarray}
\Big(\epsilon+\frac{r^2}{l^2} f(r)\Big)''=2\left[1- \frac{l^3}{r^3} \sqrt{M} \left(\alpha_3 M+\frac{2 \epsilon \alpha_2}{3 \alpha_1}\right)\right], \label{eq:second}
\end{eqnarray}
through the second derivative, where when evaluated at $r=r_{\text{ext}}$, the above equation can be written as
\begin{eqnarray}
2\left(3-\alpha_1 \sqrt{M} \frac{l}{r_{\text{ext}}}\right).
\end{eqnarray}
For Case 1, we note that when eq. (\ref{eq:second}) is zero, an inflection point arises (denoted as $r_{\text{inf}}>0$). For the special conditions
\begin{eqnarray}
&&\Big(\epsilon+\frac{r^2}{l^2} f(r)\Big)'{\Big{|}}_{r=r_{\text{inf}}}=\left(\frac{9 \sqrt{M} (3\alpha_3 M \alpha_1+2 \alpha_2 \epsilon)}{\alpha_1}\right)^{\frac{1}{3}}\nonumber\\
&&-\alpha_1 \sqrt{M}<0,\label{eq:in1}\\
&&\left(3-\alpha_1 \sqrt{M} \frac{l}{r_{\text{ext}^{(1)}}}\right)<0, \qquad  \left(3-\alpha_1 \sqrt{M} \frac{l}{r_{\text{ext}^{(2)}}}\right)>0,\nonumber
\end{eqnarray}
the existence of two extreme values is ensured, as depicted in Figure \ref{fig2}. 
\begin{figure}[h!]
  \centering
    \includegraphics[scale=0.22]{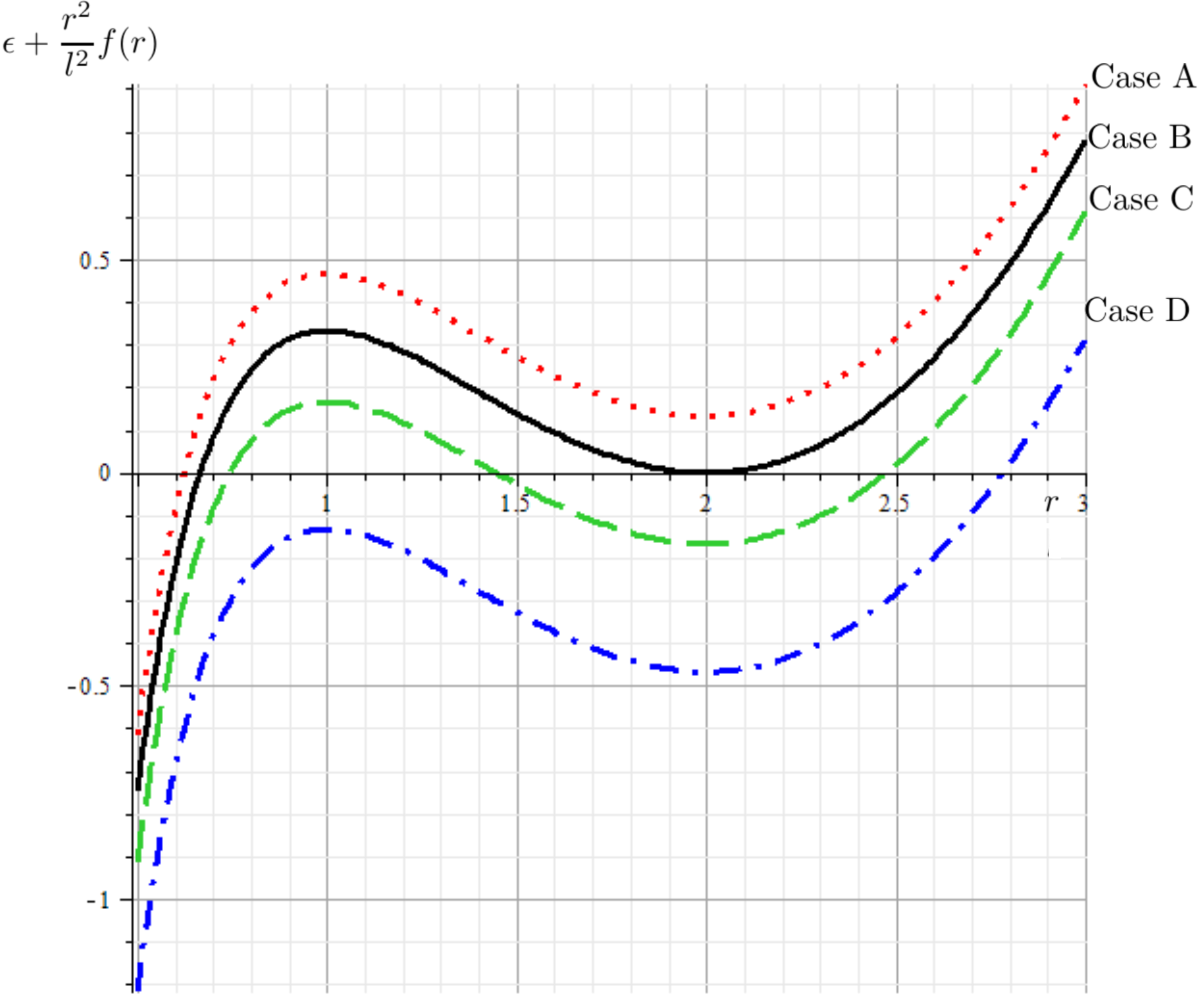}
  \caption{\label{fig3} Behavior of  $\left(\epsilon+\frac{r^2}{l^2} f(r)\right)$ for Case 1 when the inequalities (\ref{eq:in1}) are satisfied. In this context, Case A (Case D), given by the red dotted line (blue dashed-dotted line), represents the situation when the expression (\ref{eq:fext}) is positive (negative) for all $i$ (one horizon). Case B, denoted through the black curve, corresponds when  (\ref{eq:fext}) is positive at $r_{\text{ext}^{(1)}}$, while that at $r_{\text{ext}^{(2)}}$ is zero for (\ref{eq:ext}) and (\ref{eq:fext})  (extremal situation). Case C represents where three horizons exist. Here, (\ref{eq:fext}) is positive at $r_{\text{ext}^{(1)}}$, but becomes negative when $r=r_{\text{ext}^{(2)}}$, as depicted by the green dashed curve.  }
\end{figure}

Under this scenario,  $r_{\text{ext}^{(1)}}$ corresponds to a local maximum, while $r_{\text{ext}^{(2)}}$ represents a local minimum. It is important to note that $r_{\text{ext}^{(1)}}<r_{\text{inf}}<r_{\text{ext}^{(2)}}$, and the determination of whether there are one or three horizons is contingent upon the sign of
\begin{eqnarray}
\displaystyle{\Big(\epsilon+\frac{r^2}{l^2} f(r)\Big){\Big{|}}_{r=r_{\text{ext}^{(i)}}}, \mbox{ with } i=\{1,2\}.}\label{eq:fext}
\end{eqnarray}
All these situations can be represented graphically in Figure \ref{fig3}.
\begin{figure}[h!]
  \centering
    \includegraphics[scale=0.2]{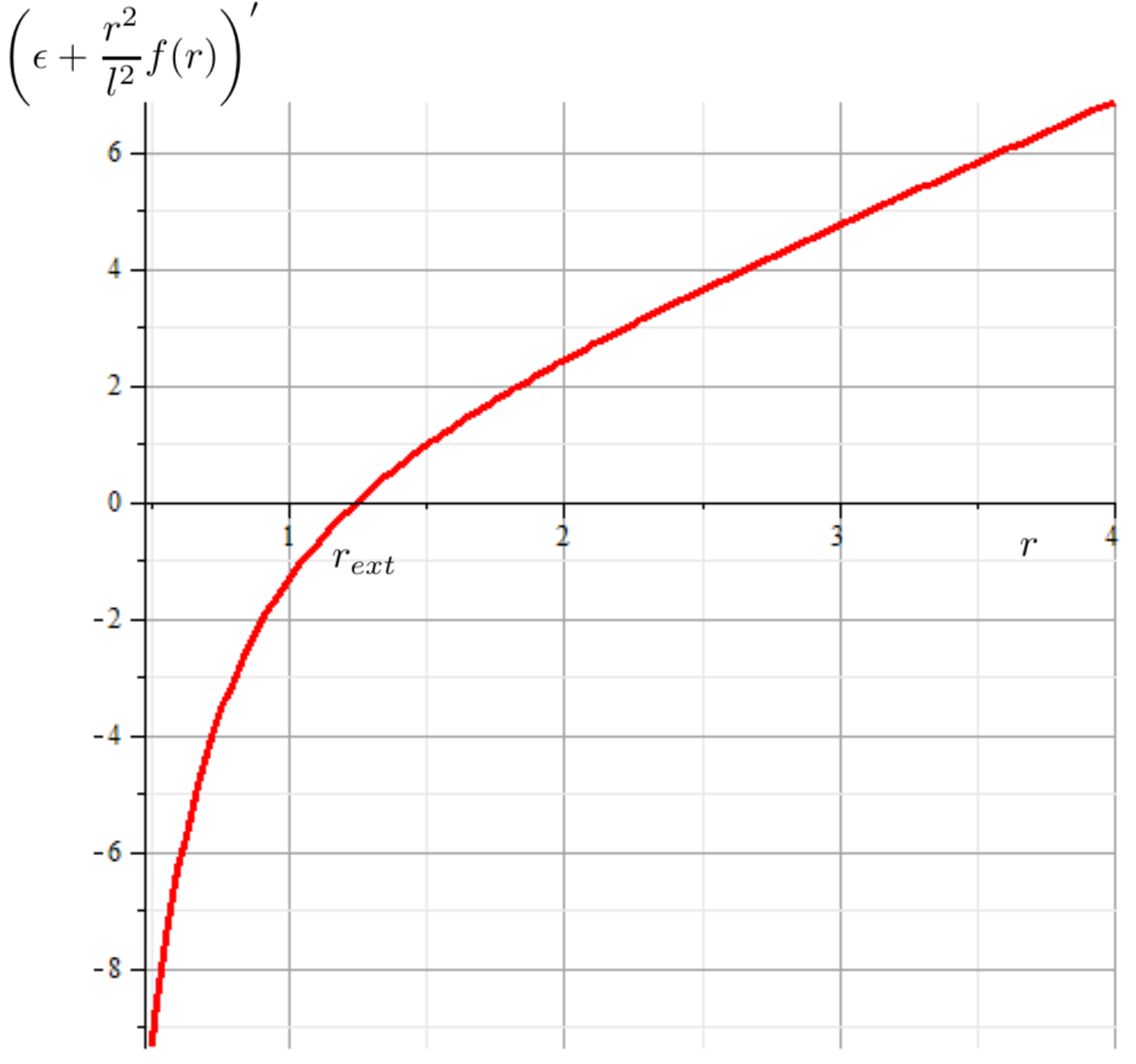}
  \caption{\label{fig4} Behavior of  $\left(\epsilon+\frac{r^2}{l^2} f(r)\right)'$ for Case 2. In this situation, we note the existence of only one extreme value  $r_{\text{ext}}>0$, and the metric function $\left(\epsilon+\frac{r^2}{l^2} f(r)\right)$ is concave up. For our calculations,  we consider  $M=1,\alpha_1=\alpha_2=1,\alpha_3=-3,\epsilon=l=1$.}
\end{figure}
\begin{figure}[h!]
  \centering
    \includegraphics[scale=0.13]{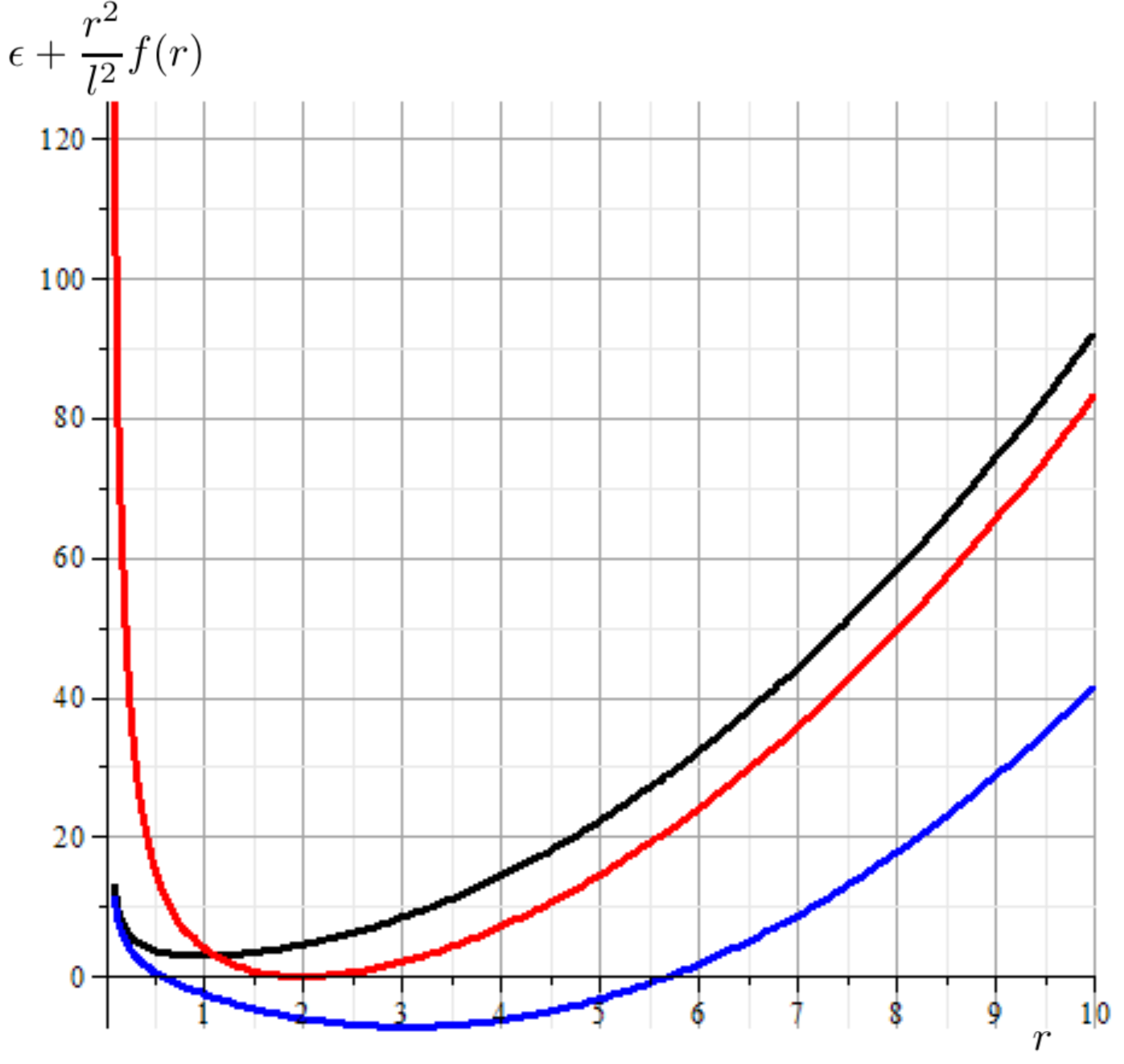}
  \caption{\label{fig5} Behavior of  $\left(\epsilon+\frac{r^2}{l^2} f(r)\right)$ for Case 2. In this situation, we note the existence of two horizons when (\ref{eq:fext}) at $r=r_{\text{ext}}$ is negative (blue curve). On the other hand, the extremal case (red curve) is when (\ref{eq:ext}) and (\ref{eq:fext}) are zero at  $r=r_{\text{ext}}$, while that there is no a presence of horizons when (\ref{eq:fext}) at $r=r_{\text{ext}}$ is positive (black curve).}
\end{figure}

In contrast, for Case 2, it can be observed from eq. (\ref{eq:second}) that there are no inflection points for $r>0$. Additionally, examining its first derivative from eq. (\ref{eq:ext}), we note the existence of only one extreme value, as depicted in Figure \ref{fig4}. This implies that $\epsilon+{r^2 f(r)}/{l^2}$ behaves without any change in concavity and the existence of horizons depend on the sign of (\ref{eq:fext}) at $r=r_{\text{ext}}$, as it shown in Figure \ref{fig5}.

Building upon the analysis and framework established for this new four-dimensional charged BH configuration, and considering the interplay between the integration constant $M$ and the coupling constants $\alpha_i$'s, the subsequent section will delve the thermodynamic quantities.

%%%%%%%%%%%%%%%%%%%%%%%%%%%%%%%%%%%%%%%%%%%%%%%%%%%%%%%
\section{Thermodynamic properties}\label{Sec-Termo}
%%%%%%%%%%%%%%%%%%%%%%%%%%%%%%%%%%%%%%%%%%%%%%%%%%%%%%%

\begin{figure}[h!]
  \centering
    \includegraphics[scale=0.12]{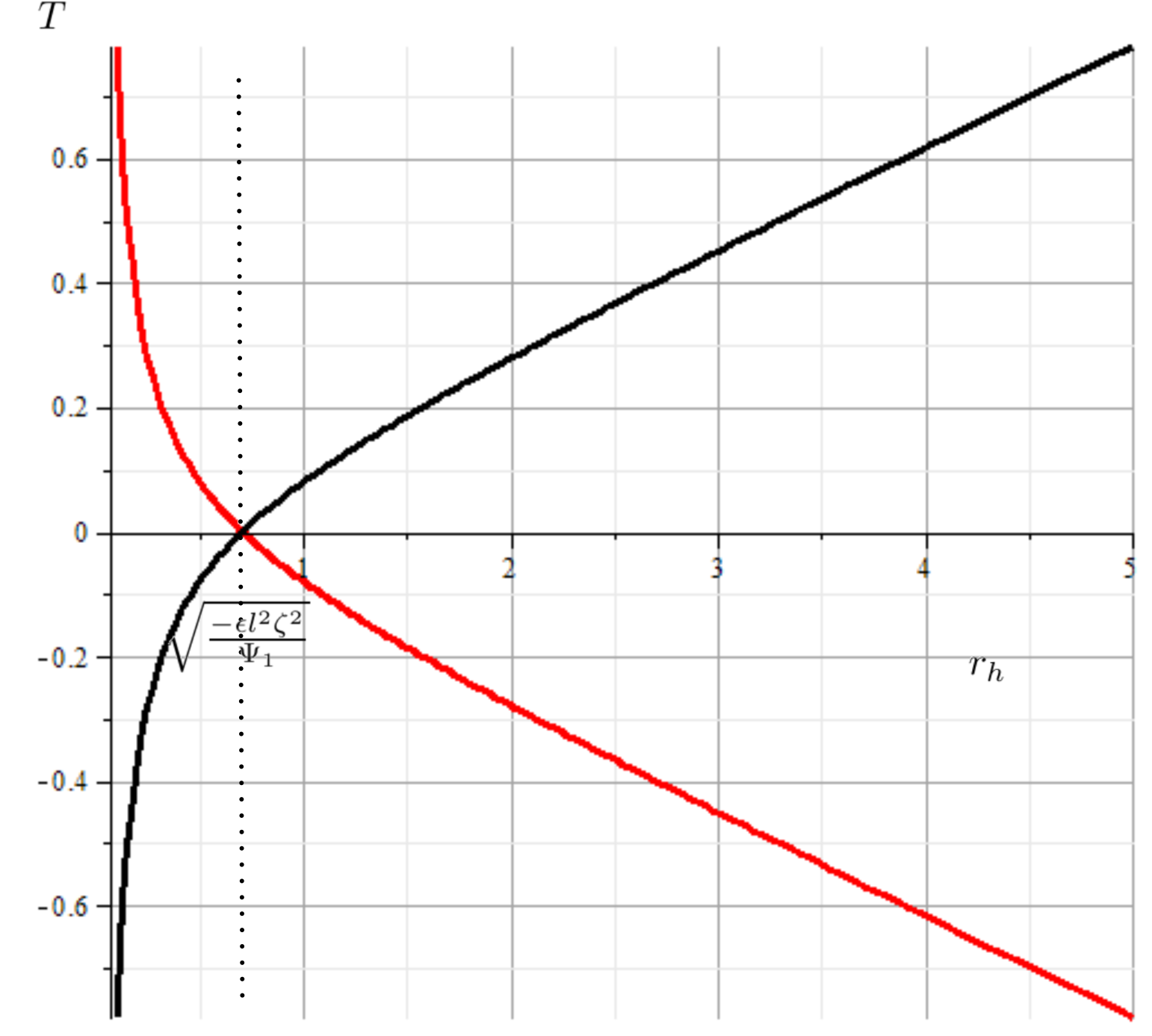}
  \caption{\label{fig5a} The behavior of $T$ in the function of the location of the event horizon $r_h$. In this situation, we note that for $\Psi_1>0$ and $\epsilon=-1$ $T$ is an increasing function, where when $r_h > \sqrt{{-\epsilon l^2 \zeta ^2}/{\Psi_1}}$ we have a positive temperature (black curve). On the other hand, when $\Psi_1<0$ and $\epsilon=1$, $T$ is a decreasing function, where $T>0$ for $0<r_h<\sqrt{{-\epsilon l^ 2 \zeta^2}/{\Psi_1}}$ (red curve).}
\end{figure}

In this section, we derive the thermodynamic quantities of the charged solution (\ref{eq:metric4d})-(\ref{eq:Lambda}), and to simplify our calculations, we consider 
\begin{equation}\label{eq:rh}
r_h=\zeta \sqrt{M} l,
\end{equation} 
with $\zeta$ is a positive constant. Among these quantities, the first one of interest is the Hawking Temperature (denoted as $T$), which can be calculated as follows:
\begin{eqnarray}
T&=&\frac{1}{4\pi} \Big(\epsilon+\frac{r^2}{l^2} f(r)\Big)' \Big{|}_{r=r_h}\nonumber\\
&=&\frac { 3 r_{h}}{4 \pi l^{2} }-\frac {\alpha_{1}\sqrt {M}
}{2 \pi l }+\frac {\alpha_{2}M}{4 \pi r_h}+\frac {\epsilon}{
4 \pi r_{h}}\nonumber\\
&=& \frac{r_h \Psi_1}{ 4 \pi \zeta^2 l^{2}} +\frac {\epsilon}{
4 \pi r_{h}}, \label{eq:T}
\end{eqnarray}
where 
\begin{eqnarray}\label{eq:psi1}
\Psi_1=3\zeta^2-2 \alpha_1 \zeta+\alpha_2.
\end{eqnarray}
Here, $r_h$ represents the position of the event horizon and for some special condition, we obtain $T>0$. Indeed, when $\epsilon=0$ (the planar case), the Hawking temperature is positive when $\Psi_1>0$, where occurs when $\alpha_1>0$ and $3\alpha_2-\alpha_1^2>0$. On the other hand, for $\epsilon \neq 0$, via the derivative of $T$ with respect to $r_h$, we have that:
$$\frac{dT}{dr_h}= \frac{\Psi_1}{ 4 \pi \zeta^2  l^{2}} -\frac {\epsilon}{
4 \pi r_h^2},\quad \frac{d^2T}{dr_h^2}=\frac{\epsilon}{2 \pi r_h^3},$$
where $T$ reaches a local extremum at
\begin{eqnarray}\label{eq:extr}
r_h^{*}=\zeta l \sqrt{\frac{\epsilon}{\Psi_1}},
\end{eqnarray}
and the concavity depends on the sign of $\epsilon$. 
To obtain a real $r_h^{*}$, we can to separate the cases for both spherical ($\epsilon=1$) and hyperbolic situation ($\epsilon=-1$). For $\epsilon=1$ we have that
$$ \Psi_1>0, \quad \frac{d^2T}{dr_h^2}>0, \quad T^{*}=T\big{|}_{r=r_h^{*}}=\frac{\sqrt{\epsilon \Psi_1}}{2 \pi l \zeta}>0,$$
ensuring positivity for $T$. For $\epsilon=-1$,  $r_{h}^{*}$ is real when $\Psi_1<0$ and a local minimum is obtained. Nevertheless, it is important to note that here $T<0$. For the remaining situation, $T$ is an increasing (for $\Psi_1>0$ and $\epsilon=-1$) or decreasing function (for $\Psi_1<0$ and $\epsilon=1$) on $r_h$, as shown in Figure \ref{fig5a}.

Additionally, the remaining thermodynamic parameters can be determined through the Wald formalism \cite{Wald:1993nt, Iyer:1994ys}. The first step in this formalism involves considering the variation of the action (\ref{eq:totaction}):
\begin{eqnarray}
&&\delta S=\mathcal{E}_{\mu \nu} \delta g^{\mu \nu} +\mathcal{E}^{\nu}_{F} \delta A_{\nu}+\mathcal{E}^{\mu \nu}_{P} \delta  \mathcal{P}_{\mu \nu} +\partial_{\mu} {\cal{J}}^{\mu}. \label{deltaS}
\end{eqnarray}
For our notations,  $\mathcal{E}^{\nu}_{F}$, $\mathcal{E}^{\mu \nu}_{P}$  and $\mathcal{E}_{\mu \nu} $ represent the equations of motions
(\ref{eq:Maxwell}), (\ref{eq:constitutive}) and (\ref{eq:Einstein}) respectively, while that  ${\cal{J}}^{\mu}$ is a surface term, given by
\begin{eqnarray}
{\cal{J}}^{\mu}&=&\sqrt{-g}\Bigg[2\left(P^{\mu (\alpha\beta)
\gamma}\nabla_{\gamma}\delta g_{\alpha\beta}-\delta g_{\alpha\beta} \nabla_{\gamma}P^{\mu(\alpha\beta)\gamma}\right) \nonumber\\
&+&\frac{\delta \mathcal{L}_{\tiny{\mbox{NLE}}}}{\delta ( \partial_{\mu} A_{\nu})} \delta
A_{\nu}
\Bigg] \label{eq:surface} ,
\end{eqnarray}
where
\begin{eqnarray}
P^{\alpha \beta \gamma \delta}&=&\frac{\delta \mathcal{L}_{\tiny{\mbox{CG}}}}{\delta
R_{\alpha \beta \gamma \delta}}=\frac{1}{4 \kappa}\,\Big(g^{\alpha\gamma}g^{\beta\delta}-g^{\alpha\delta}g^{\beta\gamma}\Big)
\nonumber\\
&-&\frac{1}{4 \kappa \Lambda}\, R \left(g^{\alpha \gamma } g^{\beta
\delta } -g^{\alpha \delta } g^{\beta \gamma
}\right)\nonumber\\
&+&\frac{3}{8  \kappa \Lambda}\, \big(g^{\beta \delta } R^{\alpha
\gamma }-g^{\beta \gamma } R^{\alpha \delta } -g^{\alpha \delta }
R^{\beta \gamma }\nonumber\\
&+&g^{\alpha \gamma } R^{\beta \delta
}\big) , \nonumber\\
\frac{\delta \cal{L}_{\tiny{\mbox{NLE}}}}{\delta (\partial_{\mu} A_{\nu})}&=&-\mathcal{P}^{\mu \nu}.\label{eq:P,A}
\end{eqnarray}
Through the surface term (\ref{eq:surface}), we define a $1$-form ${\cal{J}}_{(1)}={\cal{J}}_{\mu} dx^{\mu}$ and  its Hodge dual ${\Theta}_{(3)}=-(*{\cal{J}}_{(1)})$. After employing the equations of motion ($\mathcal{E}_{\mu \nu}$, $\mathcal{E}^{\nu}_{F}$, and $\mathcal{E}^{\mu \nu}_{P}$), we obtain 
\begin{equation}
{\cal{J}}_{(3)}={\Theta}_{(3)}-i_{\xi}*(\mathcal{L}_{\tiny{\mbox{CG}}}+\mathcal{L}_{\tiny{\mbox{NLE}}})=d(* {\cal{J}}_{(2)}).\label{eq:relwald}
\end{equation}
Here, $i_{\xi}$ denotes a contraction of the vector field $\xi^{\mu}$ on the first index of $*(\mathcal{L}_{\tiny{\mbox{CG}}}+\mathcal{L}_{\tiny{\mbox{NLE}}})$. The equation (\ref{eq:relwald}) allows to define a $2-$form  ${Q}_{(2)}=*{\cal{J}}_{(2)}$ such that ${\cal{J}}_{(3)}=d Q_{(4)}$, given by
\begin{eqnarray}
Q_{(2)}&:=& Q_{\alpha_1 \alpha_2}=\epsilon_{\alpha_1 \alpha_2 \mu \nu} Q^{\mu \nu}\label{eq:noether} \\
&=&\epsilon_{\alpha_1 \alpha_2 \mu \nu} \Big[2P^{\mu\nu\rho\sigma}\nabla_\rho \xi_\sigma -4\xi_\sigma
\nabla_\rho P^{\mu\nu\rho\sigma}\nonumber \\
&-&\frac{\delta \cal{L}}{\delta ( \partial_{\mu} A_{\nu})} \xi^{\sigma} A_{\sigma}\Big], \nonumber
\end{eqnarray}
with $\epsilon_{\alpha_1 \alpha_2 \alpha_3 \alpha_4}$ representing the Levi-Civita tensor. Subsequently, we consider the vector field $\xi^{\mu}$ as a time-translation vector, representing a Killing vector that becomes null at the event horizon location (denoted as $r_h$ as before). 

Taking all these elements into account, we can now express the variation of the Hamiltonian $\delta \mathcal{H}$ in the following form:
\begin{eqnarray}
\delta \mathcal{H}&=&\delta \int_{\mathcal{C}} {\cal{J}}_{(3)} -\int_{\mathcal{C}} d \left(i_{\xi} \Theta_{(3)}\right) \nonumber \\
&=& \int_{\Sigma^{(3)}}\left(\delta {Q}_{(2)}-i_{\xi} {\Theta}_{(3)}\right),\label{var_diff}
\end{eqnarray}
with
\begin{eqnarray}
&&\delta {Q}_{(2)}-i_{\xi} {\Theta}_{(3)}=\left[-\frac{r^2 l^2 f}{6 \kappa} \delta (f''')-\frac{r l^2 (2 f- f' r)}{12 \kappa} \right. \nonumber\\
&&\times \delta (f'')-\frac{l^2 ( f' r-2f)}{6 \kappa} \delta (f')\label{eq:dH1} \nonumber\\
&&\left.+\frac{(r^3l^2 f'''+4l^2f-2l^2 r f'+12 r^2)}{12 r \kappa} \delta f  \right] \Omega_{2,\epsilon}\nonumber\\
&&-\Omega_{2,\epsilon} A\, \delta \left(r^{2} A' \left(\frac{\partial \mathcal{H}}{\partial \mathcal{P}}\right)^{-1}\right)', 
\end{eqnarray}
where  $\Omega_{2,\epsilon} $ is the finite volume of the compact  base manifold, $\mathcal{C}$ represents a Cauchy Surface with boundary $\Sigma^{(2)}$. According to the Wald formalism, the variation (\ref{var_diff}) can be decomposed into two main components: one located at infinity $ \mathcal{H}_{\infty}$ and the other at the horizon $\mathcal{H}_{+}$, and the first law of black hole thermodynamics is a consequence of
\begin{eqnarray}
\delta \mathcal{H}_{\infty}=\delta \mathcal{M}=T\delta \mathcal{S}+\Phi_{e} \delta \mathcal{Q}_{e}=\delta \mathcal{H}_{+}.\label{eq:deltaH}
\end{eqnarray}
Here, $\mathcal{M}$, $\mathcal{S}$, and $\mathcal{Q}{e}$ represent the mass, entropy, and electric charge. Meanwhile, $T$ and $\Phi_{e}$ take the role of temperature and electric potential, respectively.  For our situation, given the solution (\ref{eq:metric4d})-(\ref{eq:Lambda}), we have that at the infinity 
\begin{eqnarray}\label{eq:dHpinf}
 \delta {\cal{H}}_{\infty} &=&\left(\frac{\alpha_1 \alpha_2  r_h^2 \Omega_{2,\epsilon}  } {3 \kappa l^2 \zeta^2}\right) \delta r_h,
\end{eqnarray}
while that at the horizon
\begin{eqnarray}\label{eq:dHrh}
\delta {\cal{H}}_{+} &=&T\Omega_{2,\epsilon} \left( \frac{4 \pi r_h \alpha_1}{\zeta \kappa}-\frac{8 \pi r_h \alpha_2}{3 \zeta^2 \kappa}+\epsilon \Gamma\right)\delta r_h\nonumber\\
&+&\Phi_e \Omega_{2,\epsilon}\left(\frac{2 r_h}{l^2 \zeta^2}\right) \delta r_h,
\end{eqnarray}
where, as before, $T$ is the Hawking temperature (\ref{eq:T}), the electric potential $\Phi_{e}$ is defined as
\begin{eqnarray}\label{eq:phie}
\Phi_{e}&=&-A_{t}(r)\big{|}_{r=r_h}\nonumber\\
&=&\left(-\frac{\alpha_1 r^2}{2 \sqrt{M} l \kappa}+\frac{\alpha_2 r}{\kappa}
-\frac{M l^2 (3 \alpha_3 \alpha_1-\alpha_2^2)}{3 \kappa r}\right)\Big{|}_{r=r_h},\nonumber\\
&=&\frac{r_h }{\kappa}\left(\frac{\alpha_1 \alpha_2}{6 \zeta}- \frac{\Psi_1 \Psi_2}{6 \zeta^2}\right) -\frac{\epsilon l^2 \Psi_2}{3 r_h \kappa},
\end{eqnarray}
where
\begin{eqnarray}
\Psi_2&=&3 \alpha_1 \zeta-2 \alpha_2,\label{eq:psi23}
\end{eqnarray}
while that
$\Gamma$ represents the contribution arising from the event horizon's topology, its expression can be written as follows:
\begin{eqnarray*}
&&\Gamma=-\frac{8 \pi l^2}{3 \kappa r_h}+\frac{4 \zeta l^2 (2 \epsilon \zeta l^2+6 r_h^2 \zeta-r_h^2 \alpha_1) \pi}{3 r_h \kappa (3 \zeta^2 r_h^2+\alpha_2 r_h^2-2 \alpha_1 \zeta r_h^2+\epsilon \zeta^2 l^2)}.
\end{eqnarray*}
Finally, identifying (\ref{eq:deltaH}), (\ref{eq:dHpinf}) and (\ref{eq:dHrh}), the mass $\mathcal{M}$, entropy $\mathcal{S}$ and electric charge $\mathcal{Q}_{e}$ read
\begin{eqnarray}
\mathcal{M}&=&\frac{\alpha_1 \alpha_2  r_h^3\Omega_{2,\epsilon}  } {9 \kappa l^2 \zeta^2}, \label{eq:mass}\\
\mathcal{S}&=& \frac{2 \pi \Psi_2 \Omega_{2,\epsilon}}{3 \kappa \zeta^2} \left[ {r_h^2}+\frac{\epsilon \zeta^2 l^2}{\Psi_1} \ln\left(\Psi_{1} r_h^2+\epsilon \zeta^2l^2 \right)\right],\label{eq:entropy}\\
\mathcal{Q}_{e}&=&\frac{r^2_h \Omega_{2,\epsilon}}{l^2 \zeta^2}\label{eq:charge},
\end{eqnarray}
where the $\Psi_{i}$'s are given in (\ref{eq:psi1}) and (\ref{eq:psi23}) respectively.

\begin{figure}[h!]
  \centering
    \includegraphics[scale=0.15]{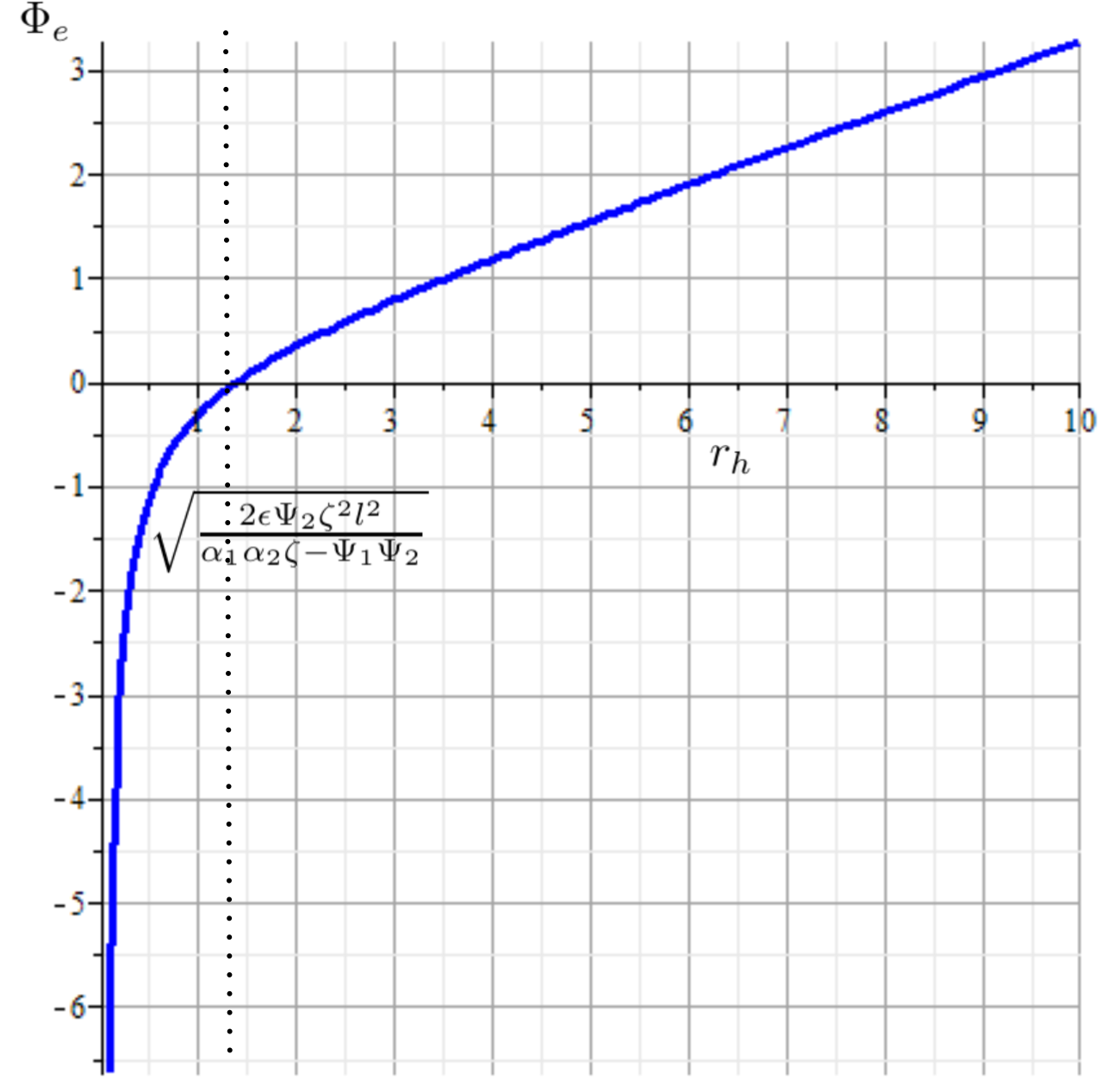}
  \caption{\label{fig5b} The behavior of $\Phi_e$ in the function of the location of the event horizon $r_h$. In this situation, we note that for $\epsilon\Psi_2>0$ together with $\alpha_2 \alpha_1 \zeta-\Psi_1 \Psi_2>0$, the electrical potential behaves as an increasing function, where $\Phi_{e}>0$ for $r_h>\sqrt{{(2\epsilon \Psi_2\zeta^2 l^2)}/{(\alpha_1 \alpha_2 \zeta-\Psi_1 \Psi_2})}$.}
\end{figure}

Let us notice that, analogous to the Hawking temperature $T$, for some suitable election of the constants $\alpha_i$'s and $\zeta$, the thermodynamic parameters (\ref{eq:phie}) and (\ref{eq:mass})-(\ref{eq:charge}) are positive. Firstly, the charge $\mathcal{Q}_{e}$ (\ref{eq:charge}) is always a positive quantity, due to that the principal components are the event horizon $r_h>0$, the  AdS radius $l>0$, the finite volume element of the compact base manifold $\Omega_{2,\epsilon}>0$, and from eq. (\ref{eq:rh}) $\zeta>0$, while that $\mathcal{M}>0$ only if $\alpha_1 \alpha_2 >0$. For the case of the entropy $\mathcal{S}$, the presence of the topology of the event horizon (represented by $\epsilon$) yields a logarithmic behavior,
where for $\Psi_1>0$, $\Psi_2>0$ as well as $\Psi_{1} r_h^2+\epsilon \zeta^2l^2>1$  we ensured that $\mathcal{S}>0$. Additionally, for $\Phi_{e}>0$, we consider:
\begin{eqnarray*}
\frac{d\Phi_e}{dr_h}=\frac{1}{\kappa} \left(\frac{\alpha_1 \alpha_2}{6 \zeta}- \frac{\Psi_1 \Psi_2}{6 \zeta^2}\right) +\frac{\epsilon l^2 \Psi_2}{3 r^2_h \kappa},\,\,
\frac{d^2\Phi_e}{dr_h^2}=-\frac{\epsilon l^2 \Psi_2}{3 \kappa r_h^3},
\end{eqnarray*}
and the concavity depends on the sign of $-\epsilon\Psi_2$. Here, the electric potential $\Phi_{e}$ reaches an extremum at
$r_h^{*}=\sqrt{-(2 \epsilon \zeta ^2 l^2 \Psi_2)/(\alpha_2 \alpha_1 \zeta-\Psi_1 \Psi_2)},$ where when $\epsilon\Psi_2<0$ and $\alpha_2 \alpha_1 \zeta-\Psi_1 \Psi_2>0$ we have that 
$$\frac{d^2\Phi_e}{dr_h^2}>0,\quad \Phi_{e}\big{|}_{r=r_h^{*}}=\sqrt{-\frac{2 \epsilon l^2 \Psi_2 (\alpha_1\alpha_2 \zeta-\Psi_1 \Psi_2)}{9 \kappa^2 \zeta^2}}>0,$$
ensuring that $\Phi_{e}>0$. On the other hand,  when $\epsilon\Psi_2>0$ and $\alpha_2 \alpha_1 \zeta-\Psi_1 \Psi_2>0$, the electrical potential behaves as an increasing function, as shown in Figure \ref{fig5b}.

\begin{figure}[h!]
  \centering
    \includegraphics[scale=0.25]{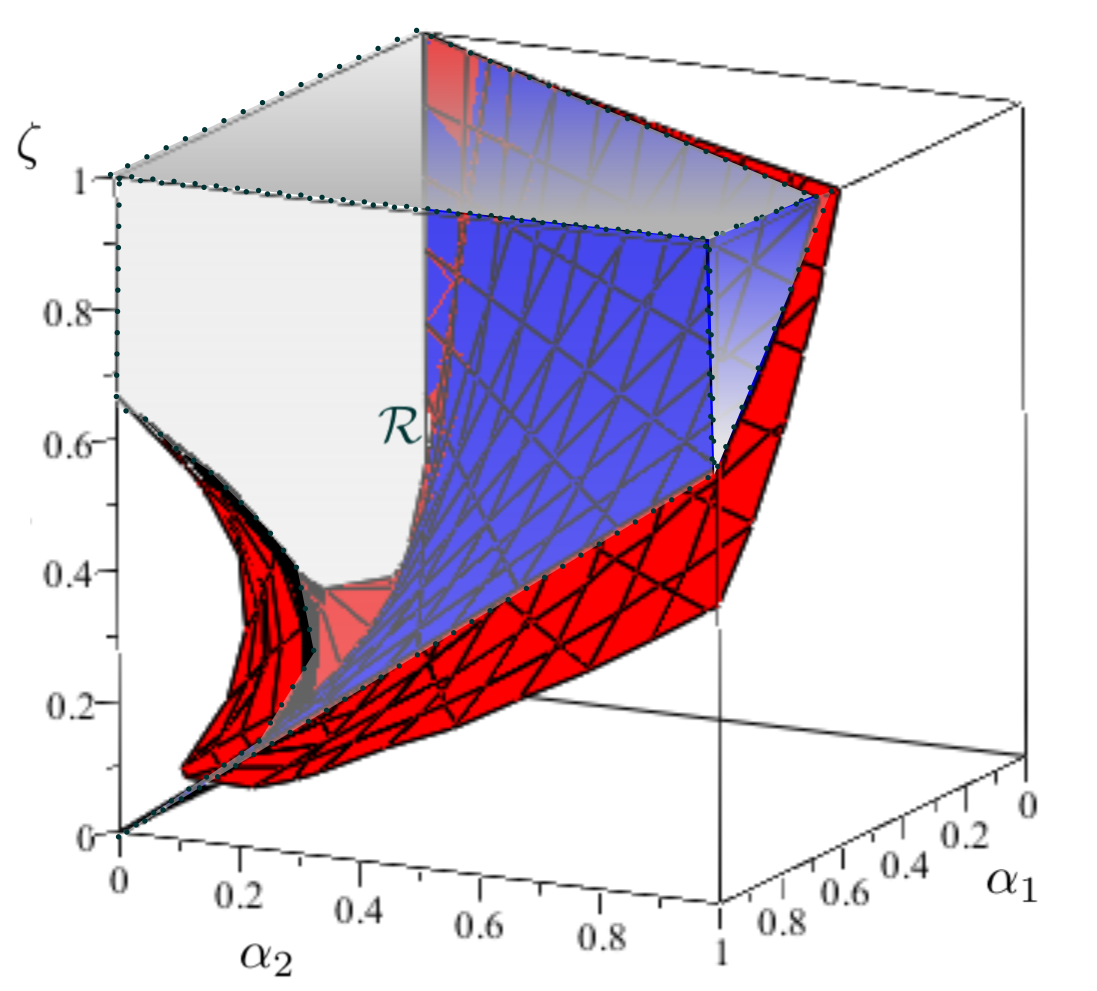}
  \caption{\label{fig6} The region $\mathcal{R}$ represents the stable local solution under electrical and thermal fluctuations. In this context, for $\alpha_1>0$ and $\alpha_2>0$, the blue surface denotes $\Psi_2=0$, the black surface corresponds to $\Psi_1=0$, and the red surface represents $\alpha_2 \alpha_1 \zeta + \Psi_2 \Psi_1=0$.}
\end{figure}

\begin{figure}[h!]
  \centering
    \includegraphics[scale=0.28]{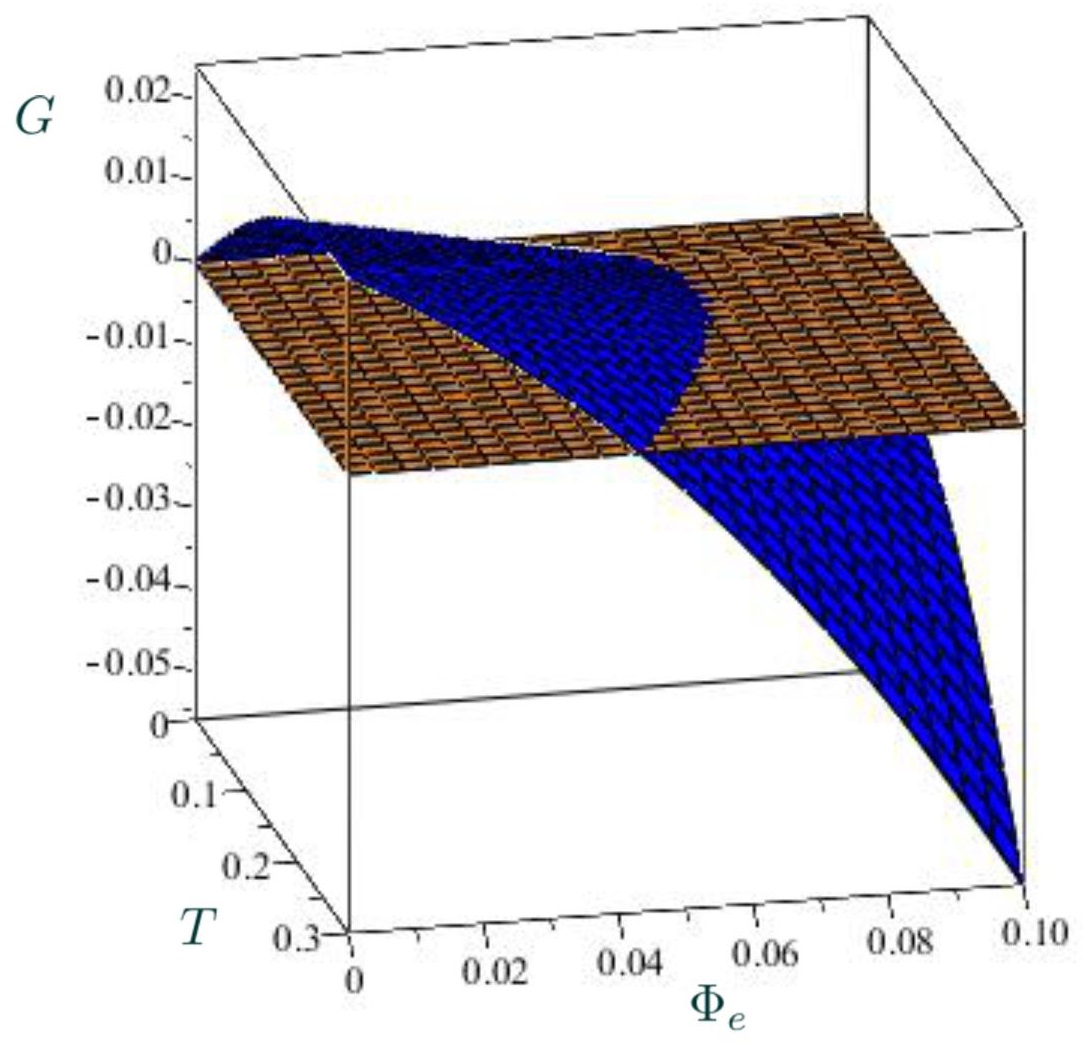} 
    \includegraphics[scale=0.25]{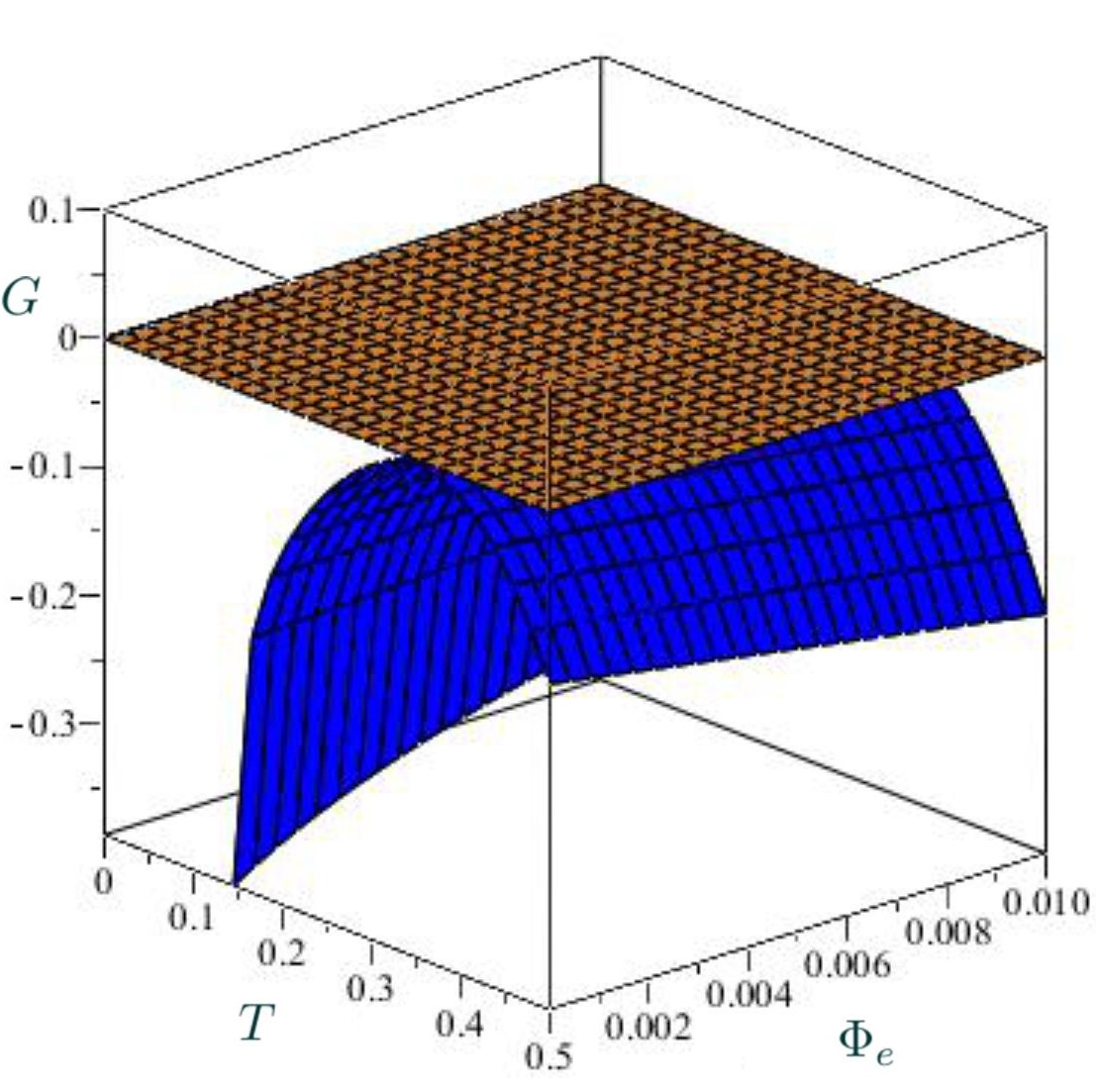}
\caption{\label{fig6a-b}Up Panel: Gibbs free energy $G$ for $\epsilon=1$ and $\zeta=0.7$. Right Panel: $ G$ for the situation $\epsilon=-1$ together with $\zeta=0.8$. For both cases, the gold surface represents the reference $G=0$ and for our simulations, the remainder constants are fixed to unity.}
\end{figure}

On the other hand, we can examine the system's response to small perturbations around equilibrium by considering these thermodynamic quantities. To achieve this, the extensive thermodynamic quantities $\mathcal{M}$ and $\mathcal{Q}_{e}$ can be expressed as functions of the intensive ones $T$ and $\Phi_e$. From eqs. (\ref{eq:T})-(\ref{eq:psi1}) and (\ref{eq:phie})-(\ref{eq:psi23}), we note that the location of the event horizon $r_h$ can be expressed in the following way
\begin{eqnarray}\label{eq:rh}
r_h=F(T,\Phi_e)=\frac{2\zeta^2 (4 \pi l^2 \Psi_2 T + 3 \kappa \Phi_{e})}{ (\alpha_2 \alpha_1 \zeta+\Psi_2 \Psi_1 )},
\end{eqnarray}
which is positive for $\Psi_2>0$,$\Psi_1>0$ and $\alpha_1 \alpha_2>0$. Now, the mass $\mathcal{M}$ as well as the electric charge $\mathcal{Q}_{e}$ take the form
\begin{eqnarray*}
\mathcal{M} (T,\Phi_{e})&=& \frac{\alpha_1 \alpha_2   \Omega_{2,\epsilon} F^3(T,\Phi_e) }{9 \kappa \zeta^3 l^2},\\
\mathcal{Q}_{e} (T,\Phi_e)&=&\frac{\Omega_{2,\epsilon}  F^2(T,\Phi_e)}{l^2 \zeta^2},
\end{eqnarray*}
and with this, we are in a position to analyze the local thermodynamic (in)stability under thermal and electrical fluctuations, represented via the specific heat $C_{\Phi_{e}}$ and the electric permittivity $\epsilon_{T}$, given by
\begin{eqnarray}\label{eq:cesp-elp}
C_{\Phi_{e}}&=&\left(\frac{\partial \mathcal{M}}{\partial T}\right)_{\Phi_{e}}=\frac{\alpha_1 \alpha_2 \Omega_{2,\epsilon} F^2}{3 \kappa \zeta^3 l^2} \left( \frac{\partial F}{\partial{T}}\right)\nonumber\\
&=&\frac{ 8 \pi \alpha_1 \alpha_2 \Psi_2  \Omega_{2,\epsilon} F^2 }{3 \zeta \kappa (\alpha_2 \alpha_1 \zeta+\Psi_2 \Psi_1)},\label{eq:cesp} \\
\epsilon_{T}&=&\left(\frac{\partial \mathcal{Q}_{e}}{\partial \Phi_{e}}\right)_{T}=\frac{2  \Omega_{2,\epsilon} F }{l^2 \zeta^2} \left(\frac{\partial F}
{\partial{\Phi_{e}}}\right) \nonumber\\
&=&\frac{12 \kappa \Omega_{2,\epsilon} F }{l^2  (\alpha_2 \alpha_1 \zeta+\Psi_2 \Psi_1)},\nonumber
\end{eqnarray}
where the sub-index stands for constant electric potential $\Phi_{e}$ and constant temperature $T$ respectively. It is important to note that the non-negativity of equations given in (\ref{eq:cesp})  ensure local stability under thermal fluctuations ($C_{\Phi_{e}} \geq 0$) and electrical fluctuations ($\epsilon_{T} \geq 0$). For example, for $\alpha_1 \alpha_2>0$, $\alpha_2 \alpha_1 \zeta+\Psi_2 \Psi_1>0$, $\Psi_1 >0$, and $\Psi_2 \geq 0$ we have that $C_{\Phi_{e}} \geq 0$ and $\epsilon_{T} \geq 0$, as shown in Figure \ref{fig6}. Together with the above, with eq. (\ref{eq:rh}) and in the grand canonical ensemble where the temperature and electric potential are fixed, we can compute the Gibbs free energy ${G}(T,\Phi_e)=\mathcal{M}-T \mathcal{S}-\Phi_{e} \mathcal{Q}_{e}$: 
\begin{eqnarray}\label{eq:Gibbs}
G&=&\frac{\alpha_1 \alpha_2  \Omega_{2,\epsilon} F^3 }{9 \kappa \zeta^3 l^2}-\frac{2 \pi \Psi_2 \Omega_{2,\epsilon} T F^2}{3 \kappa \zeta^2}-\frac{\Phi_{e}  F^2 \Omega_{2,\epsilon}}{l^2 \zeta^2}\nonumber\\
&-&\frac{2 \pi \Omega_{2,\epsilon} \epsilon  l^2 \Psi_2 T}{3 \kappa  \Psi_1} \ln\big(\Psi_{1} F^2+\epsilon \zeta^2l^2 \big).
\end{eqnarray}
Although the previous equation is a priori unwieldy, via some simulations, we can obtain interesting cases when the coupling constants $\alpha_i$'s are fixed. For the spherical situation, we note a first-order PT  for some special relation between $T$ and $\mathcal{Q}_{e}$ (see Fig. \ref{fig6a-b}, up panel). For the hyperbolic case,  we note that ${G}<0$, and the charged BH has lower free energy than the thermal AdS space-time, being the preferred state (see Fig. \ref{fig6a-b}, down panel).

%%%%%%%%%%%%%%%%%%%%%%%%%%%%%%%%%%%%%%%%%%%%%%%%%%%%%%%%%%%%%%%%%%%%%%%%%
\section{Quasinormal modes and greybody factors calculations}\label{Sec-QNM}
%%%%%%%%%%%%%%%%%%%%%%%%%%%%%%%%%%%%%%%%%%%%%%%%%%%%%%%%%%%%%%%%%%%%%%%%%

The evolution of small perturbations in BHs generally goes through three stages: the initial outburst, the QNM ringing, and the final power law tail. Here, the QNM ringing stage corresponds to the BH's QNM, which typically exhibits damped oscillations in the form of a discrete set of complex frequencies, where the real part represents the frequency of the oscillation and the imaginary part represents the rate at which the oscillation decays. As was shown previously in the introduction, the QNMs are characterized by being independent of the initial perturbation and only dependent on the BH parameters, making them an important tool for studying BHs and gravity theories.

In the following lines, we will divide the content into four parts. The initial three segments elucidate various calculation methodologies, the first one delineates the numerical calculation of the dynamic evolution of a scalar field through the finite element method, then in the second part we elaborate on computing the QNMs across the complex plane using pseudospectral methods.  For the third part, we introduce the concept of the greybody factors, and finally, we dedicate it to the presentation and analysis of the calculation results.

%%%%%%%%%%%%%%%%%%%%%%%%%%%%%%%%%%%%%%%%%%%%%%%%%%%%%%%%%%%
\subsection{The Finite Element Method}\label{Sec-A}
%%%%%%%%%%%%%%%%%%%%%%%%%%%%%%%%%%%%%%%%%%%%%%%%%%%%%%%%%%%

For this situation, we chose $\epsilon=1$  and then, $d\Omega_{2,1}^2$ from equation (\ref{eq:dOmega2}) corresponds to the metric of a $2-$ dimensional unit sphere. Defining now 
\begin{eqnarray}\label{eq:F}
F(r)=1+\frac{r^2 f(r)}{l^2}, 
\end{eqnarray}
as the metric function, the line element  (\ref{eq:metric4d}) becomes to:
\begin{eqnarray}\label{eq:metric4d_F}
	ds^2=-F(r)dt^2+\frac{dr^2}{F(r)} +r^2 d\Omega_{2,1}^2,
\end{eqnarray}
and the fluctuations of a massless scalar field $\Psi$ can be described via the Klein-Gordon (KG) equation
	\begin{eqnarray}\label{eq:KG-equation}
		\frac{1}{\sqrt{-g}} \partial_{\mu}\left(\sqrt{-g} g^{\mu \nu} \partial_{\nu} \Psi\right)=0.
	\end{eqnarray}
Here, we assume that the coordinates $\{t,r,\theta,\rho\}$ of the scalar field $\Psi$ are separated in the form 
$$\Psi(t,r,\theta,\rho)=\Phi(t,r)Y_{lm}(\theta,\rho),$$ 
where $Y_{lm}(\theta,\rho)$ represents a spherical harmonic function. Substituting $\Psi$ into eq. (\ref{eq:KG-equation}) yields the following differential equation
    \begin{eqnarray}\label{eq:Separation}
	    -\frac{{\partial^2 \Phi}}{{\partial t^2}} + \frac{{\partial^2 \Phi}}{\partial r_{*}^2} - V(r)\Phi = 0.
	\end{eqnarray}
	Assuming that $\Phi=e^{-i \omega t} \psi$ and substituting it into eq. (\ref{eq:Separation}), $t$ and $r$ can be further separated, resulting in:
	\begin{eqnarray}\label{eq:inear equation about psi}
	    \frac{{\partial^2 \psi}}{{\partial r_{*}^2}} + \left(\omega^2 - V(r)\right)\psi = 0.
	\end{eqnarray}
Here, ${r_*}$ represents the tortoise coordinate, which is defined as $d{r_*}={dr}/{dF(r)}$, $\omega$ is the frequency of QNM and $V(r)$ represents the effective potential, where for spherically symmetric cases
    \begin{eqnarray}
    	V(r) = F(r)\left[\frac{L(L+1)}{r^2} + \frac{1}{r} \left(\frac{dF(r)}{dr}\right) \right],
    \end{eqnarray}
where $L$ is the angular quantum number. 

The equation (\ref{eq:Separation}) can be numerically integrated using a finite element method to obtain the evolution plot in the time domain. First, rewriting equation  in the form of a difference equation yields:
	\begin{eqnarray}
		\Psi_{h+1,k}&=&\left[2 - \frac{2(\Delta t^2)}{(\Delta r_*)^2} - \Delta t^2 V_k\right] \Psi_{h,k}\notag\\
        &-&\Psi_{h-1,k} +\frac{(\Delta t^2)}{(\Delta r_*)^2} \big(\Psi_{h,k-1} + \Psi_{h,k+1}\big),
	\end{eqnarray}
	where $\Psi_{h,k} = \Psi(h\Delta t, k\Delta r_*)$, $V_k = V(k\Delta r_*)$. In this case, $r_*$ ranges from negative infinity to a constant value denoted as $r_{*\tiny{\mbox{infinite}}}$, representing the spatial infinity. Perturbations in this case do not propagate indefinitely towards both sides as they do in asymptotically flat spacetime, but rather behave like vibrations in a half-infinite vibrating string. In the asymptotic AdS spacetime, the commonly used boundary conditions are Dirichlet and Neumann conditions. Here, we use the Dirichlet condition $\Psi(t,r_*=r_{*\tiny{\mbox{infinite}}}) = 0$ and the initial conditions are
	\begin{eqnarray}
		&&\Psi(t=0,r_*)=C_1 \exp\left(-C_2 (r_* - C_3)^2\right),\\
		&&\left.\frac{\partial  \Psi\left(t,r_{*}\right)}{\partial t}\right|_{t=0}=0,
    \end{eqnarray}
	where $C_1$, $C_2$, $C_3$ are constants for adjusting the initial perturbation. In order to satisfy the stability condition of the scalar field, we choose $\Delta t/(\Delta r_*) = 2/3$ (see Ref. \cite{Lin:2016wci}).

%%%%%%%%%%%%%%%%%%%%%%%%%%%%%%%%%%%%%
\subsection{The pseudospectral method for computing QNMs.}\label{Sec-B}
%%%%%%%%%%%%%%%%%%%%%%%%%%%%%%%%%%%%%

There are three main methods for calculating QNMs of AdS black holes. The first one is the series solution \cite{Horowitz:1999jd}, which  requires initial frequency guessing and iterative numerical calculations to obtain the results, thus requiring a significant amount of trial and error to find the actual frequencies. The second one is the continued fraction method \cite{Nollert:1993zz,Berti:2004um}, which was utilized to compute QNMs of Schwarzschild-AdS black holes \cite{Daghigh:2022uws}. The third one, which we will employ in the present work, is the pseudospectral method \cite{Jansen:2017oag}, allowing us to calculate the frequencies of massless scalar field QNMs, via the Mathematica package \cite{Jansen:2017oag}. The approach of this method involves discretizing the quasi-normal equation using spectral methods, solving the resulting generalized eigenvalue equation directly. The specific procedure is as follows, to calculate the QNMs, we need to solve for the complex frequency $\omega$ as given in eq. (\ref{eq:inear equation about psi}). In order to facilitate the use of the pseudospectral method, we first transform the coordinates to Eddington-Finkelstein coordinates, rewriting the metric (\ref{eq:metric4d_F}) as
    \begin{eqnarray}
    	ds^2 = -F(z)dv^2 - \left(\frac{2}{z^2}\right)dzdv + \left(\frac{1}{z^2}\right) d\Omega_{2,1}^2,
    \end{eqnarray}
    where $z=1/r$, $v=t+r_*$. Without loss of generality, assuming that the event horizon (or outer) is located at $r_h=1$, we can obtain that $z=1$ ($z=0$) corresponds to the event horizon (spatial infinity). Therefore, we have that $z \in \left(0,1 \right]$. By substituting the scalar field $\Psi=e^{-i \omega t} Y_{lm}(\theta,\rho)\psi(z)$ under Eddington-Finkelstein coordinates into the KG equation (\ref{eq:KG-equation}), we obtain \cite{Jansen:2017oag}:
    \begin{eqnarray}\label{eq:radial equation}
    	&&\left[zL(L+1)+2i\omega\right]\psi - \left(2iz\omega + z^3 F'(z)\right)\psi' \\
    	&&-z^3 F(z)\psi''= 0, \nonumber
    \end{eqnarray}
where now $(')$ represents the derivative with respect to $z$ for a function.

When considering boundary conditions, near the event horizon, only ingoing solutions are permissible, while near infinity, only outgoing solutions exist. In the case of AdS BHs, a method to fulfill these boundary conditions involves redefining
\begin{eqnarray}\label{eq:psit}
 \psi(z)=z^2\tilde{\psi}(z),
\end{eqnarray}
ensuring that the normalizable solutions remain continuous both at the event horizon and the boundary. Meanwhile, the non-normalizable solutions tend to diverge and oscillate rapidly. The above allows us to select solutions that satisfy the boundary conditions. For instance, near the event horizon, the ingoing solution remains smooth, whereas the outgoing solution oscillates rapidly. Conversely, at infinity, the outgoing solution tends toward zero, while the ingoing solution diverges. 

Regarding the calculation of QNM frequency, there is another detail: the previous context assumes the event horizon $r_h=1$. However, when arbitrary numerical values are substituted into the parameters of $F(z)$, the resulting event horizon may not necessarily be located at $r_h=1$. Therefore, a simple adjustment is required here. We denote the event horizon obtained after substituting arbitrary numerical values into $F$ parameters as $r_h'$, which provides a natural length scale, allowing us to non-dimensionalize our variables. That means $l$ is expressed in units of $r_h'$ and the QNM frequency $\omega$ is in units of $r_h'^{-1}$. Therefore, we need to perform straightforward replacements for $l$ in $F(z)$ and $\omega$ in eq.(\ref{eq:radial equation}):
\begin{eqnarray}\label{eq:replacement}
	l\rightarrow l/r_h', \qquad \omega \rightarrow \omega r_h',
\end{eqnarray}
such that it can guarantee the event horizon is located at $rh=1$.
Taking into account eq.(\ref{eq:replacement}) and substituting eq.(\ref{eq:psit}) into eq.(\ref{eq:radial equation}), we obtain the final form of the QNM equation, which reads:
 \begin{eqnarray}\label{eq:QNM equation}
    	&\left[2i\omega r_h' + zL(L+1) - 2z^2 i\omega r_h'- 2z^3 F - 2z^4 F'\right] \tilde{\psi}  \notag\\
    	&-\left(2z^3 i\omega r_h'+ 4z^4 F + z^5 F'\right) \tilde{\psi}' -z^5 F(z) \tilde{\psi}''= 0.
    \end{eqnarray}
To numerically solve the above equation, it is necessary to discretize the equation by replacing the continuous variables with a collection of discrete points, also known as collection points, and the set of points is denoted a grid. There are different ways to choose the discrete points, in this work we adopt the Chebyshev grid. Here, any arbitrary function can be approximately represented as a sum of products of basis functions and corresponding coefficients, where the coefficients represent the values of the function at Chebyshev nodes. The basis functions $C_J(z)$ are linear combinations of Chebyshev polynomials, whose specific forms can be found in \cite{Jansen:2017oag}. For our situation, we consider $\tilde{\psi}(z)$ as
   \begin{eqnarray}
    	&&\tilde{\psi}(z) \approx \sum_{J=0}^{N} \tilde{\psi}(z_J) C_J(z),\\
    	&&z_J = \cos\left(\frac{J}{N} \pi \right), \quad J=0,\ldots,N, \quad N \in \mathbb{N},
    \end{eqnarray}
where $z_J$ represents the Chebyshev grid points. Since the QNM equation involves first and second derivatives of $\tilde{\psi}$ with respect to $z$, it is necessary to take derivatives of $C_J(z)$. Let's denote the derivative matrix as $D_{IJ}^{(1)} = C_I'(z_J)$ and the second derivative matrix as $D_{IJ}^{(2)} = C_I''(z_J)$. Now, the problem of solving the QNM equation can be transformed into solving a matrix equation, allowing us to rewrite eq. (\ref{eq:QNM equation}) in the following form:
\begin{eqnarray}
    	&&c_{0,0}(z) = zL(L+1) - 2z^3F - 2z^4F',\notag\\
    	&&c_{0,1}(z) = 2i - 4z^2i,\notag\\
    	&&c_{1,0}(z) = -(4z^4F + z^5F'),\notag\\
    	&&c_{1,1}(z) = -(2z^3i),\notag\\
    	&&c_{2,0}(z) = -z^5F(z),\notag\\
    	&&c_{2,1}(z) = 0,
    \end{eqnarray}
together with
    \begin{eqnarray}
   &&\left[c_{0,0}(z) + \omega c_{0,1}(z)\right] \tilde{\psi} + \left[c_{1,0}(z) + \omega c_{1,1}(z)\right] \tilde{\psi}' \nonumber\\
&&+ [c_{2,0}(z) + \omega c_{2,1}(z)] \tilde{\psi}'' = 0,
    \end{eqnarray}
    which can also be written in matrix form
    \begin{eqnarray}\label{eq:matrix equation}
    	(M_0 + \omega M_1) \tilde{\psi} = 0.
    \end{eqnarray}
Here, $(M_0)_{IJ} = c_{0,0}(z_I) \delta_{IJ} + c_{1,0}(z_I) D_{IJ}^{(1)} + c_{2,0}(z) D_{IJ}^{(2)}$  and $\delta_{IJ}$ is the Kronecker function. The definition of $(M_1)_{IJ}$ is similar. Finally, eq.(\ref{eq:matrix equation}) can be solved directly or by inputting the QNM eq. (\ref{eq:QNM equation}) into the program package in \cite{Jansen:2017oag} for computation.

%%%%%%%%%%%%%%%%%%%%%%%%%%%%%%%%%
 \subsection{Calculation of greybody factors}
%%%%%%%%%%%%%%%%%%%%%%%%%%%%%%%%%

When considering the quantum effects of a BH, it can emit Hawking radiation at its event horizon. However, due to the gravitational effects near to it (acting as a barrier),  we observe that is not a direct radiation spectrum, but rather a greybody spectrum that has been influenced by the curvature of spacetime, describing via the greybody factor $\tau_{\tiny{\mbox{WKB}}}$. This factor represents the transmission coefficient of Hawking radiation and can be obtained from the gravitational potential of the BH. 

In the following lines, we will study this using the Wentzel–Kramers–Brillouin (WKB) approximation and rigorous bound methods, following the steps performed in \cite{Boonserm:2023oyt,Visser:1998ke}. It is worth pointed out that the WKB method is convenient for studying cases where the effective potential is a barrier, while the rigorous bound method can only provide a lower bound for the greybody factor. Although it may not give the exact values, it is still helpful for qualitative analysis. 

Considering the WKB approximation, we can approximate $\omega^2\approx V_{\tiny{\mbox{max}}}$ for eq.(\ref{eq:inear equation about psi}) , where $V_{\tiny{\mbox{max}}}$ is the maximum value of the effective potential near the barrier. The expression for the greybody factor is \cite{Cho:2004wj}
    \begin{eqnarray}
    	\tau_{\tiny{\mbox{WKB}}}(\omega) = \frac{1}{1 + \exp\left[{{\pi (V_{\tiny{\mbox{max}}} - \omega^2)}/{\sqrt{-V''(r_{*{\tiny{\mbox{m}}}})/2}}}\right]},
    \end{eqnarray}
where $r_{*{\tiny{\mbox{m}}}}$ means $V(r_{*{\tiny{\mbox{m}}}})=V_{\tiny{\mbox{max}}}$. 

The general semi-analytic bounds for the greybody factors are given by \cite{Boonserm:2023oyt,Visser:1998ke}
    \begin{eqnarray}\label{eq:greybound}
    	\tau_b(\omega) \geq \mbox{sech}^2\left[\int_{-\infty}^{+\infty}  \varXi \,dr_{*}\right].
    \end{eqnarray}
with 
$$\varXi =\frac{\sqrt{(h)'^2+(\omega^2-V-h^2)^2}}{2h}.$$ 
Here, $h=h(r_*)$ is a function which must satisfy two conditions. The first one is that is a positive function ($h>0$) and the second one is 
$$h(-\infty)=\lim_{r_{*}\rightarrow -\infty} h(r_{*})=\lim_{r_{*}\rightarrow +\infty} h(r_{*})=h(+\infty)=\omega.$$ 
Without loss of generality, we can set $h=\omega$ and the equation (\ref{eq:greybound}) can be simplified as
    \begin{eqnarray}
    	\tau_b(\omega) \geq \mbox{sech}^2\left[\int_{r_h}^{r_e} \frac{|V(r)|}{|F(r)|} \,dr\right],
    \end{eqnarray}
where $r_e$ means the value of the coordinate $r$ at the local minimum point of the effective potential (this is, $V(r_e) \approx 0$).

%%%%%%%%%%%%%%%%%%%%%%%%%%%%%%%%%%%%%%%%
\subsection{The calculation results}\label{Sec-cal}
%%%%%%%%%%%%%%%%%%%%%%%%%%%%%%%%%%%%%%%%

Initially, for the computation of QNMs the identification of the event horizon is required. Hence, this part initiates by delineating the distinct scenarios for Case 1 and Case 2  outlined previously in (\ref{eq:Cases}).

From Figure \ref{fig3} (Case 1), we can see that for Cases A and D $F(r)$ has only one zero point, which suggests it as the event horizon. For Case D, we note that there is no potential barrier outside the event horizon, indicating the absence of QNMs and allowing us to exclude it. In Case B, due to $F(r)$ is positive between the two zero points, we consider the first zero point as the event horizon and the second one as the cosmological horizon. In Case C, between the second and third zero points, $F(r)$ is negative, representing a one-way membrane region of the BH. Thus, we consider the third zero point as the event horizon. However, similarly to Case D, there is no potential barrier outside the event horizon, indicating the absence of QNMs. On the other hand, from Figure \ref{fig5} (Case 2), when $F(r)$ has no zero points, it implies the naked singularity of the BH, which violates the strong cosmic censorship hypothesis and we do not consider this scenario. Among the remaining cases, neither of them exhibits QNMs outside the event horizon, except for Cases A and B, where we will focus on the first one.

\begin{figure}[h!]
	\includegraphics[scale=0.4]{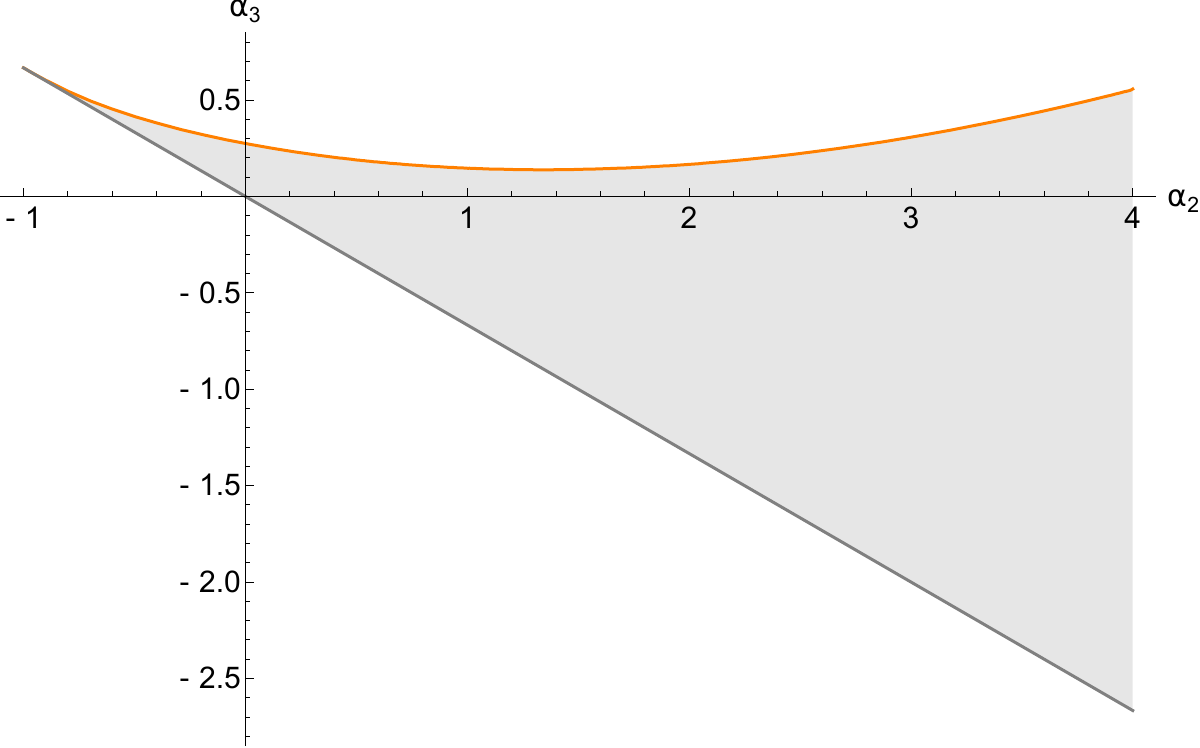}
	\caption{For Case A, we have plotted a parameter range for $\alpha_{2}$ and $\alpha_{3}$ corresponding to the gray-shaded region. Here, $\alpha_{1}=1,M=1, l=100,L=1$.}
	\label{fig parameter}
\end{figure}

Concerning Case A, we explore how the coupling constants $\alpha_{2}$ and $\alpha_{3}$ affect the QNMs, while keeping $\alpha_{1}$ fixed. For simplicity, in our calculations, we set the remaining parameters as $\alpha_{1}=1, M=1, l=100, L=1$. In order to better illustrate the changes in the parameter space, we first plot a partial parameter space (represented by the gray region) that satisfies Case A  in Figure \ref{fig parameter}.
\begin{figure}[h!]
	\centering
	\includegraphics[scale=0.55]{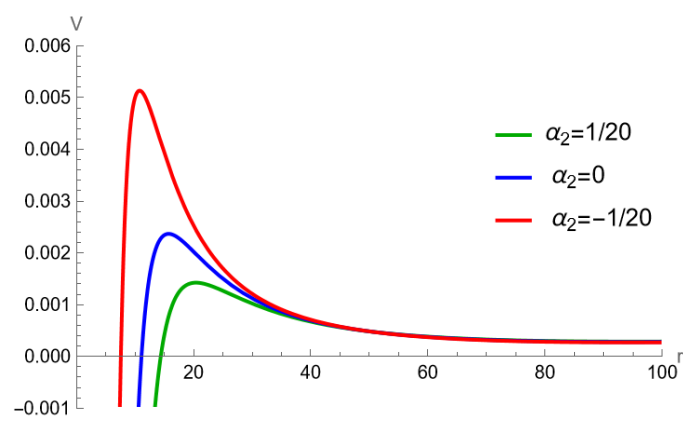}
	\caption{The effective potential curves as the function of $r$ for different values of $\alpha_{2}$ are shown. The remaining physical parameters are chosen as $\alpha_{1}=1, \alpha_{3}=1/10, M=1, l=100,L=1$. }
	\label{fig6a}
	\includegraphics[scale=0.55]{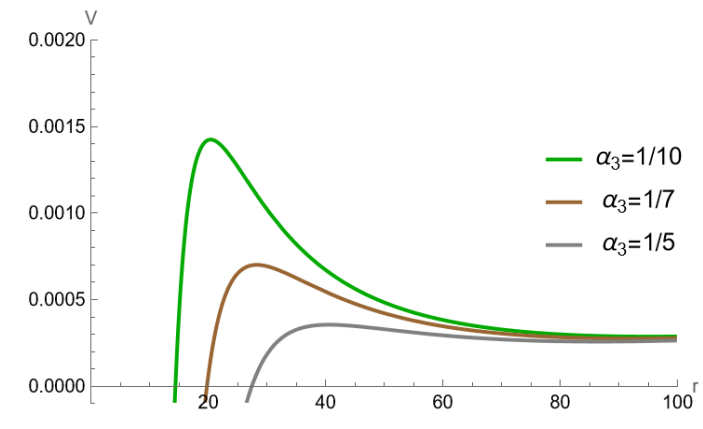}
	\caption{The effective potential curves as the function of $r$ for different values of $\alpha_{3}$ are shown. Here, we consider $ \alpha_{1}=1, \alpha_{2}=1/20,M=1, l=100,L=1$.}
	\label{fig7}
\end{figure}
\begin{figure}[h!]
     	\centering
     	\includegraphics[scale=0.6]{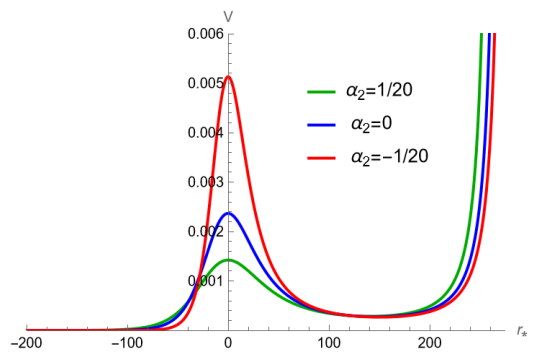}
     	\caption{The effective potential curves as the function of $r_*$ corresponding to different values of $\alpha_2$ are shown. Here, $\alpha_{1}=1, \alpha_{3}=1/10, M=1, l=100,L=1.$ }
     	\label{fig8}
     	\includegraphics[scale=0.52]{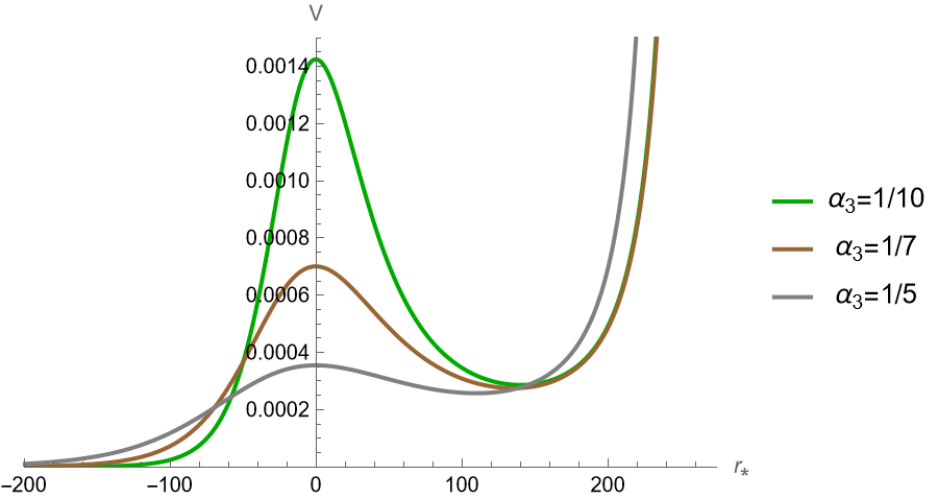}
     	\caption{The effective potential curves as the function of $r_*$ for different values of $\alpha_{3}$ are shown. Here, $\alpha_{1}=1, \alpha_{2}=1/20,M=1, l=100,L=1.$}
     	\label{fig9}
\end{figure}
The entire parameter space is an open region bounded by the gray and orange curves, where $\alpha_{2}$ has no upper bound.  As was shown in (\ref{eq:Cases}), to satisfy this situation, it is required that$$\alpha_2>-\left(\frac{3}{2}\right)\alpha_{3}.$$ However, $\alpha_2$ cannot be equal to $-3/20$ , otherwise, the function $F(r)$ will not comply with the condition of Case A and even the event horizon will not exist. 

In Figure \ref{fig parameter}, the gray line represents $\alpha_{3} = -(2/3) \alpha_{2}$, while the orange curve corresponds to the critical value of $\alpha_{3}$ with a potential barrier. When $\alpha_{3}$ is greater than the orange curve, there is no potential barrier, and the effective potential $V$ is a monotonously increasing function of the radius $r$.  As $\alpha_{2}$ decreases and reaches a certain lower limit ($\alpha_{2} \approx -1$), the gray line intersects with the orange curve. However, as $\alpha_{2}$ increases, the critical value of $\alpha_{3}$, represented by the orange curve, first decreases and then gradually increases. As shown in Figures \ref{fig6a} and \ref{fig8}, when $\alpha_2$ decreases and approaches the straight line $\alpha_3=(-2/3) \alpha_{2}$, the peak of the effective potential increases, regardless of whether it is in terms of the $r$ or $r_*$ coordinates. Furthermore, as shown in Figures \ref{fig7} and \ref{fig9}, when $\alpha_3$ increases, the effective potential $V$ decreases further. This demonstrates the situation when the parameters are selected closer to the orange curve.  Observing Figures \ref{fig6a} and \ref{fig7}, it can be noticed that when the parameters are chosen to approximate the gray line from Figure  \ref{fig parameter} , the event horizon also comes closer to the point at $r=0$.  So, in the parameter space, the general trend of the effective potential variation is as follows: the closer the parameters are to approximate the gray line, the smaller the position of the event horizon, and the higher the potential barrier. Conversely, the variation occurs in the opposite direction as one approaches the orange curve.

Together with the above, from Figures \ref{fig8} and \ref{fig9}, we can note that on the right side, as $r_*$ approaches infinite space ($r_* \to r_*{\tiny{\mbox{infinite}}}$), the effective potential $V$ rapidly increases and tends towards $+\infty$. When the perturbation approaches this region, it bounces back and propagates towards the left. 

\begin{figure}[h!]
      	\includegraphics[scale=0.52]{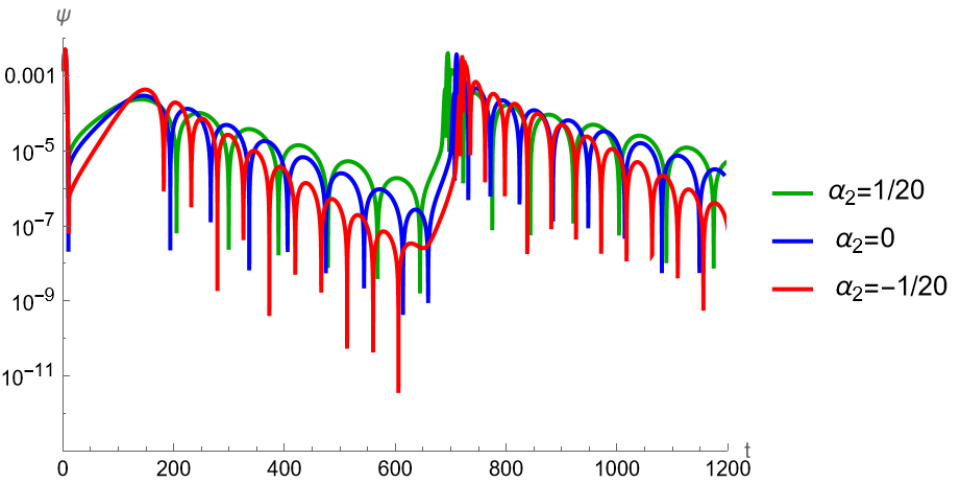}
      	\caption{The evolution of the scalar field $\psi$ in the time domain corresponding to different values of $\alpha_2$ is shown. Here, we consider $\alpha_{1}=1, \alpha_{3}=1/10, M=1, l=100,L=1$. }
      	\label{fig10}
      	\includegraphics[scale=0.52]{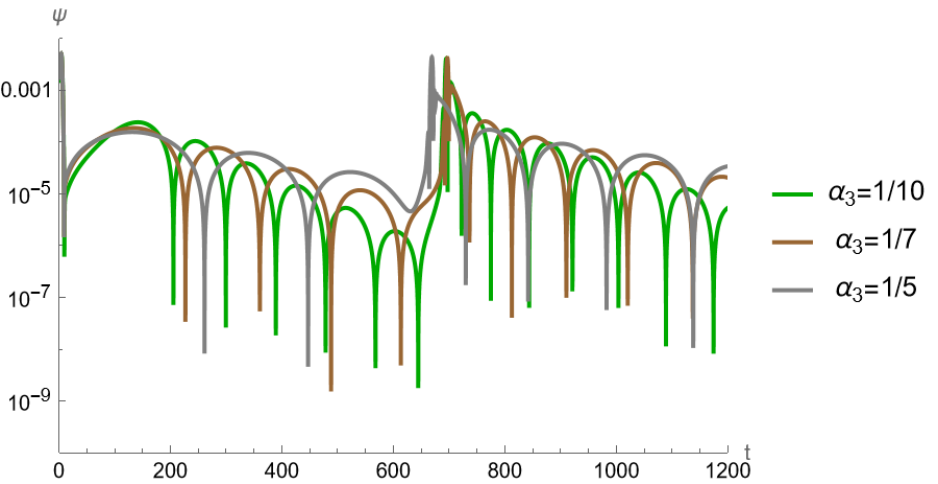}
      	\caption{The evolution of the scalar field $\psi$ in the time domain corresponding to different values of $\alpha_3$ is shown. When $\alpha_3=1/5$, the second peak of the scalar field arrives more quickly, indicating a smaller cosmic boundary and an earlier encounter with the bouncing wave. For our calculation,  $\alpha_{1}=1, \alpha_{2}=1/20, M=1, l=100$ and $L=1$ are considered.}
      	\label{fig11}
\end{figure}

We calculate the evolution of the scalar field $\psi$ in time domain $t$ by using the Finite Element Method. The initial perturbation is set on the left side of the effective potential $(r_*=-80)$, and the evolution of $\psi$ is observed to be located near the initial perturbation. From Figure \ref{fig10}, it can be seen that $\psi$ goes through an initial burst phase (approximately for $0<t<160$), followed by quasinormal ringing. As $\psi$ encounters waves reflected back from infinitely far away in space, $\psi$ increases again before cyclically going through the aforementioned process until the overall waveform gradually diminishes.  Generally, the larger the peak value of the effective potential, the larger the peak value of $\psi$ that is reached. For example, in Figure \ref{fig10}, it is noteworthy that $\psi$ corresponding to $\alpha_2=-1/20$ reaches its first local maximum for the initial burst around $t=160$ and reaches its second local maximum for the reflection waves around $t=700$. The remaining two cases of $\alpha_2$ are similar. Additionally, the oscillation frequency of $\psi$ is higher, and the decay is faster. Together with the above, from Figure \ref{fig11} we note the influence of the evolution of $\psi$ in the time domain when $\alpha_3$ is variating. Combining with the previous effective potential from Figures \ref{fig9} and \ref{fig11}, the same conclusion holds under the above findings.

 \begin{table*}[htp]
    	\caption{QNMs for different values of $\alpha_{2}$ using grids with N=200 and N=220. $(\alpha_1=1, \alpha_3=1/10, M=1, l=100, L=1)$}
      \begin{ruledtabular}
    		\begin{tabular}{llll}
    			$n$	 &$\alpha_{2}=-1/20$   &$\alpha_{2}=0$    &$\alpha_{2}=1/20$\\
    			1&	 0.02285852-$2.323 \times10^{-7}i$&   0.02296808-$2.495\times 10^{-6} i$ &0.02303847-$1.536\times 10^{-5} i$\\
    			2&	 0.07477502-0.00104947 $i$&     0.05290060-0.001302913 $i$&   0.04224606-0.001515529$i$\\
    			3&	 0.08478345-0.00206041 $i$&     0.06265967-0.002873224 $i$&   0.05187974-0.003612847$i$\\
    			4&	 0.09498565-0.00323322 $i$&     0.07274986-0.004622982 $i$&   0.06195569-0.005893477$i$\\
    			5&	 0.10533912-0.00444971 $i$&     0.08303681-0.006391512 $i$&   0.07225682-0.008179062$i$\\
    			6&	 0.11579112-0.00566995 $i$&     0.09343722-0.008150684 $i$&   0.08268312-0.010449802$i$\\
             \end{tabular}
        \end{ruledtabular}
        \label{Table:pseudospectral_QNM}
       \end{table*}

Table \ref{Table:pseudospectral_QNM} presents the results obtained using the pseudospectral method. Setting $\alpha_{3}=1/10$, we can observe that in QNMs, regardless of the value of $n$, the absolute value of the imaginary part decreases as $\alpha_{2}$ decreases. The behavior of the real part is slightly more complex. For the fundamental mode ($n=1$), as $\alpha_{2}$ decreases, the real part becomes smaller. However, for all cases except the fundamental mode ($n>1$), a decrease in $\alpha_{2}$ leads to an increase in the real part. 

\begin{figure}[h!]
	\includegraphics[scale=0.38]{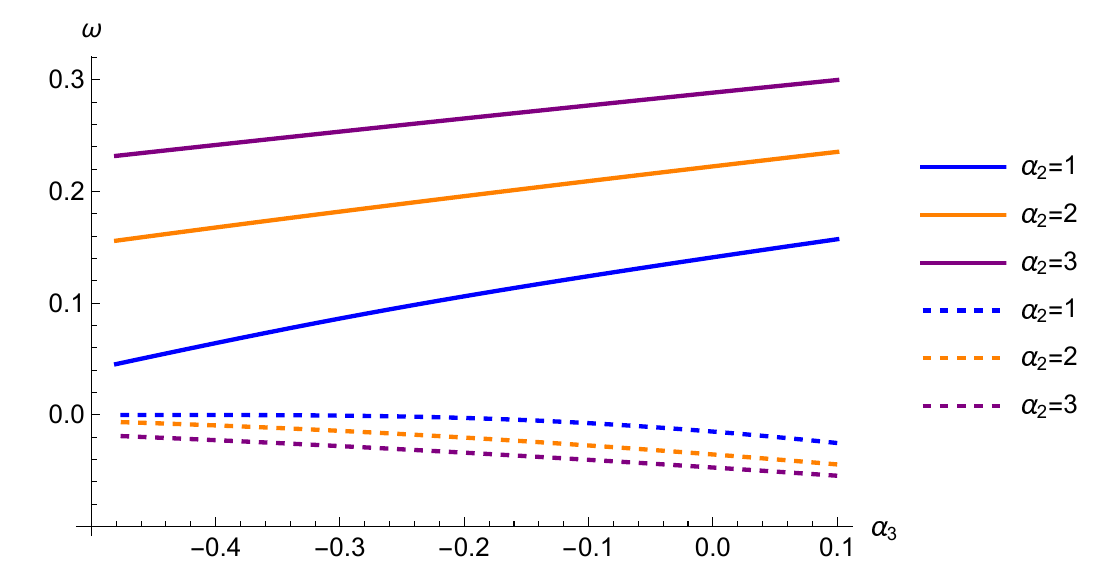}
	\caption{For $n=1$ and different values of $\alpha_{2}$, the variations of QNMs with respect to $\alpha_{3}$ have been plotted. The solid line represents the real part of the QNM, while the dashed line represents the imaginary part.}
	\label{fig qnm(n=1)alpha_3}
\end{figure} 

Now, considering the simpler case for $n=1$, we have plotted the scenarios of QNMs under two different parameter variations. In Figure \ref{fig qnm(n=1)alpha_3}, for three distinct $\alpha_{2}$ chosen, the behavior of QNMs within the interval of $\alpha_{3}$ ranging from $-0.5$ to $0.1$ exhibits a straightforward pattern. With the increase in $\alpha_{3}$, the real parts of the QNMs exhibit growth, while the imaginary parts diminish. This observation implies an escalation in the frequency of oscillations and a more rapid decay of the QNMs. 

One may wonder why the frequency computed by the pseudospectral method seems to be different from the time-domain result. Nevertheless, it should be emphasized that when calculating the time-domain evolution image using the finite element method, the waves at the cosmological boundary are not prevented from reflecting. The above can be explained as when the perturbation propagates to the cosmological boundary, it is equivalent to reaching $r_{\tiny{\mbox{*infinite}}}$ corresponding to $v \to \infty$ in Fig \ref{fig8}, at which point the perturbation encounters an infinitely high potential barrier, causing the perturbation to bounce back. It implies that perturbations cannot cross an infinitely high potential barrier. This model is analogous to a semi-infinite vibrating string. On the other hand, when calculating QNMs via the pseudospectral method, the boundary condition for $\psi$ is set to have only an outgoing solution, but no incoming solution at the boundary. It means there is no wave reflecting back. However, in the evolution diagrams from finite element calculations, perturbations bounce back due to encountering an infinitely high potential barrier. In other words, the boundary conditions between the two methods are different. Furthermore, the frequency in the time domain can also be influenced by the initial conditions. QNMs can be understood as something similar to eigenfrequencies, which may not directly correspond to the actual frequency of wave oscillations. Therefore, the results will naturally be different. 

\begin{figure}[h!]
       	\centering
       	\includegraphics[scale=0.60]{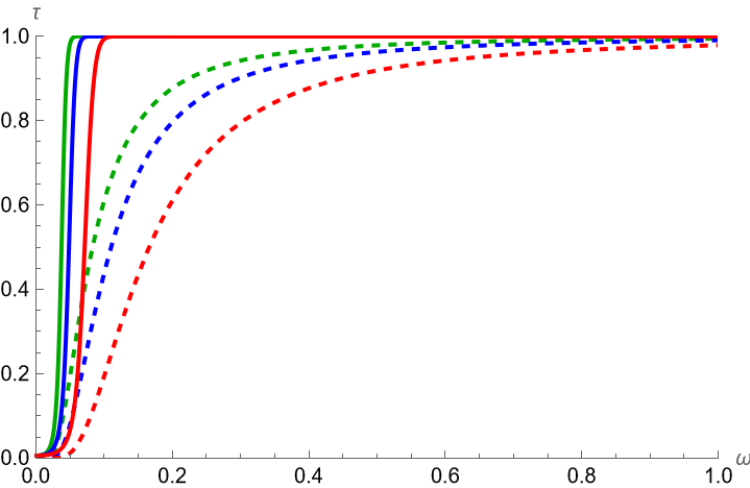}
       	\caption{The solid line corresponds to $\tau_{\tiny{\mbox{WKB}}}$, while the dashed line corresponds to $\tau_b$ and $\alpha_{2}=1/20$ (green), $\alpha_{2}=0$ (blue), $\alpha_{2}=-1/20$ (red). The remaining physical parameters are chosen as $\alpha_{1}=1, M=1, l=100,L=1.$}
       	\label{fig12}
       \end{figure}

Along with the above, from Figure \ref{fig12} we note that the outcomes derived from the rigorous lower bound consistently exhibit smaller values in comparison to those obtained through the WKB approximation. Additionally,  regardless of the results from the WKB approximation or the rigorous lower bound, we can observe that the greybody factors are lower for $\alpha_{2} = -1/20$. This point is intuitively understandable. In fact, from Figure \ref{fig8},  the effective potential $V$ exhibits nearly similar widths, where the peak of the effective potential is higher for $\alpha_{2} = -1/20$, consequently leading to a reduced transmission rate. In general, for Case A, as $\alpha_{2}$ approaches $-(3/2) \alpha_{3}$ (but not equal to $-(3/2) \alpha_{3}$), the peak value of the effective potential increases and the position of the event horizon moves closer to the point $r = 0$. Consequently, this translates into a smaller greybody factor.

%%%%%%%%%%%%%%%%%%%%%%%%%%%%%%%%%%%%%%%%
\section{Conclusions and discussions}\label{Sec-conclusions}
%%%%%%%%%%%%%%%%%%%%%%%%%%%%%%%%%%%%%%%%

In the present work, we show first that it is possible to construct four-dimensional BHs with CG \cite{Lu:2011zk} for arbitrary topology, where the matter source is characterized with a nonlinear NLEs framework, using the $(\mathcal{H,P})$ formalism, and allowing us to obtain electrically non-linear charged AdS BHs, motivated by the AdS/CFT correspondence.

It is important to note that this specific configuration is defined by three key structural constants: $\alpha_1$, $\alpha_2$, and $\alpha_3$ as well as the constant $\epsilon$, representing the topology of the event horizon, alongside the structural function $\mathcal{H}$ previously derived in \cite{Alvarez:2022upr}. These constants serve as pivotal factors in generating scenarios involving BHs characterized by one, two, or three horizons, represented via Cases 1 and 2 from (\ref{eq:Cases}), independently of the topology. 

Together with the above, the incorporation of NLE as a matter source into CG has been instrumental in obtaining non-zero thermodynamic properties, following the Wald formalism, thanks to the contribution from the coupling constants $\alpha_i$'s. It is interesting to note that the topology plays a significant role in the thermodynamic quantities. In fact, from eq. (\ref{eq:entropy}) we note that the entropy $\mathcal{S}$ enjoys logarithmic behavior. We want to reserve a crucial point to note here: As shown in Refs. \cite{Feng:2015oea,Minamitsuji:2023nvh,Bravo-Gaete:2022lno,Bravo-Gaete:2021hlc,Bravo-Gaete:2020lzs,Bravo-Gaete:2023iry} for the particular case of scalar-tensor theories, there is a difference in the entropy between the Wald formalism and the standard Wald entropy formula. The latter does not provide a complete expression for $\mathcal{S}$ that satisfies the first law (\ref{eq:deltaH}). In our case, this variation is due to the presence of the topology. As a first thermodynamic quantity, we note that the expression of the mass $\mathcal{M}$ (\ref{eq:mass}) is naturally recovered following other procedures such as the quasilocal method
\cite{Kim:2013zha,Gim:2014nba}, corresponding to an off-shell prescription of the Abbott-Desser-Tekin procedure \cite{Abbott:1981ff,Deser:2002rt,Deser:2002jk}. On the other hand, via the Wald formula the entropy takes the form
\begin{eqnarray}
\label{wald} \mathcal{S}_{W}&{=}&-2 \pi \int_{{H}} d^{2}x \sqrt{|h|} \left(P^{\mu \nu \rho \sigma} \, \varepsilon_{\mu \nu} \,  \varepsilon_{\sigma \rho}\right)\nonumber\\
&=&\frac{2 \pi \Psi_2 \Omega_{2,\epsilon} r_h^2}{3 \kappa \zeta^2}.
\end{eqnarray}
Here, the integral is evaluated on a 2-dimensional spacelike surface $H$ (denoted as the bifurcation surface), where the timelike Killing vector {$\partial_{t}=\xi^{\mu}\partial_{\mu}$} vanishes, and $|h|$ denotes the determinant of the induced metric on ${H}$, $\varepsilon_{\mu \nu}$ represents the binormal antisymmetric tensor normalized as $\varepsilon_{\mu \nu} \varepsilon^{\mu \nu}=-2$, showing us a difference between eqs. (\ref{eq:entropy}) and (\ref{wald}) given by
$$\mathcal{S}=\mathcal{S}_{W}+\frac{2 \pi \Omega_{2,\epsilon} \epsilon  l^2 \Psi_2}{3 \kappa  \Psi_1} \ln\big(\Psi_{1} r_h^2+\epsilon \zeta^2l^2 \big),$$
characterized by the presence of the topology $\epsilon$. With all the above, we note that with the temperature $T$ (\ref{eq:T}) and electric potential $\Phi_{e}$ (\ref{eq:phie}), as well as the mass $\mathcal{M}$ (\ref{eq:mass}) and the electric charge $\mathcal{Q}_{e}$ (\ref{eq:charge}), the first law (\ref{eq:deltaH}) does not fold when we consider the entropy $\mathcal{S}_{W}$ (\ref{wald}). The above deserves further investigation in their own right, being an open problem to explore.

Together with the above, this configuration exhibits local stability against both thermal and electrical fluctuations. This stability is evidenced by the non-negativity observed in both the specific heat $C_{\Phi_{e}}$ and the electric permittivity $\epsilon_{T}$, which is established under specific conditions into the model. For instance, when considering $\alpha_1 \alpha_2>0$, $\alpha_2 \alpha_1 \zeta+\Psi_2 \Psi_1>0$, alongside conditions such as $\Psi_1 >0$ and $\Psi_2 \geq 0$, where $\Psi_1$ and $\Psi_2$ are defined in (\ref{eq:psi1}) and (\ref{eq:psi23}) respectively, the local stability is ensured. Additionally, via the Gibbs free energy (\ref{eq:Gibbs}), we note that the topology of the base manifold allows us to compare this charged configuration with respect to the thermal AdS space-time, where for example with $\epsilon=1$ a first-order PT is allowed, while that when $\epsilon=-1$, the charged BH is the preferred configuration. This, from the gauge/gravity duality, is related to holographic confinement and deconfinement states \cite{Witten:1998zw}.

On the other hand, with this charged solution and the spherical situation, our second motivation is to perform numerical calculations of the dynamic evolution of a scalar field through the finite element method, and the computations of QNMs across the complex plane using pseudospectral methods \cite{Jansen:2017oag}. Here, it is important to note that the behavior of effective potential $V$ as well as the evolution of the scalar field $\psi$ in the time domain are affected when the coupling constants $\alpha_i$'s vary (see Figs. \ref{fig6a}-\ref{fig qnm(n=1)alpha_3}). 

Combining the parameter space (from Figure \ref{fig parameter}), the effective potential (from Figs. \ref{fig6a}-\ref{fig9}), and the evolution plot of the scalar field (see Figs. \ref{fig10}-\ref{fig11}), certain rules emerge. When $\alpha_{2}$ and $\alpha_{3}$ are chosen to be close to the orange curve, indicating a smaller potential barrier, the greybody increases. This corresponds to a decrease in frequency on the evolution plot, with slower decay. Conversely, when $\alpha_{2}$ and $\alpha_{3}$ are chosen to approach the gray line, signifying a larger potential barrier, the greybody decreases. This results in an increase in frequency and faster decay on the evolution plot. However, when QNMs for $n=1$ are computed using pseudospectral methods, the observed rules differ from those seen in the evolution plot. Even when approaching the orange curve, the real part increases and the imaginary part decreases. The discrepancy is attributed to the presence of two distinct boundary conditions. The QNMs exhibit a relatively straightforward change for variations in $\alpha_{3}$: the real part increases with the growth of  $\alpha_{3}$, while the imaginary part decreases with the increase in $\alpha_{3}$.

Some interesting further works include, for example, that (a) via the covariant phase space approach, we generalized the first law of black hole thermodynamics, following Refs. \cite{Gunasekaran:2012dq,Bokulic:2021dtz,Gulin:2017ycu}, where now the parameters, such as the cosmological constant $\Lambda$, play a significant role. Following this line, (b) the first law of BHs mechanics can be modified when the cosmological constant
$$P=-\frac{\Lambda}{8\pi},$$
is related to the pressure $P$ of the system \cite{Kastor:2009wy,Kubiznak:2014zwa,Kubiznak:2016qmn}, and modifying the first law as
$$\delta \mathcal{M}=T\delta \mathcal{S}+V \delta P+\Phi_{e} \delta \mathcal{Q}_{e},$$
where $V$ corresponds to the conjugate thermodynamic volume and $\mathcal{M}$ is the enthalpy \cite{Dolan:2010ha,Cvetic:2010jb}. The above has been highly analyzed from a holographic point of view \cite{Karch:2015rpa} and related to the BH chemistry.  Given that we now have the inclusion of GC as a gravity model (\ref{eq:CG}), we have to carefully consider the presence of higher-order corrections in the Wald formalism (\ref{eq:dH1}), where $\Lambda$ is present. The above has caused a generalization of Wald formalism to unveil the structure of extended BH thermodynamics, as shown in Refs.  \cite{Xiao:2023two,Hajian:2023bhq,Xiao:2023lap}. It will be interesting to study how thermodynamics results from these explorations (points (a) and (b)), which is an open problem to study.

Finally, the possibility to enrich CG with an arbitrary four-dimensional topology allows us to analyze the connection between AdS BH and quantum complexity via the gauge/gravity correspondence. Under this duality, new interesting holographic observables appear (see for example, Refs. \cite{Susskind:2014rva,Brown:2015bva,Brown:2015lvg,Couch:2016exn,Belin:2021bga,Belin:2022xmt}) where the thermodynamic quantities take a providential role.

%%%%%%%%%%%%%%%%%%%%%%%%%%%%%%%%%%%%%%%%%%%%%%%%%%%%%% 
\begin{acknowledgements}
This work is supported by National Natural Science Foundation of China (NSFC) with Grants No.12275087. M.B. is supported by PROYECTO INTERNO UCM-IN-22204, L\'INEA REGULAR. The authors thank the Referee for the commentaries and suggestions to improve the paper.
\end{acknowledgements}
%%%%%%%%%%%%%%%%%%%%%%%%%%%%%%%%%%%%%%%%%%%%%%%%%%%%%%

%%%%%%%%%%%%%%%%%%%%%%%%%%

\end{document}